\documentclass[11pt,a4paper]{article}

\usepackage{amsmath}                    

\usepackage{amssymb}                    

\usepackage{bm}                         

\usepackage{exscale,relsize}            

\usepackage{graphicx}                   

\usepackage[numbers]{natbib}            

\usepackage{stmaryrd}                   

\usepackage[caption=false]{subfig}      

\usepackage{url}                        

\usepackage{doi}                        

\urldef{\myurl}\url{http://www.math.kent.edu/~gartland}

\newcommand{\myemail}{\href{mailto:gartland@math.kent.edu}{gartland@math.kent.edu}}

\bmdefine{\ehat}{e}
\bmdefine{\nhat}{n}

\bmdefine{\nuhat}{\nu}

\bmdefine{\bmB}{B}
\bmdefine{\bmD}{D}
\bmdefine{\bmE}{E}
\bmdefine{\bmH}{H}
\bmdefine{\bmM}{M}
\bmdefine{\bmP}{P}

\bmdefine{\bmx}{x}

\bmdefine{\bfalpha}{\alpha}
\bmdefine{\bfeps}{\epsilon}
\bmdefine{\bfmu}{\mu}

\bmdefine{\bfzero}{0}

\newcommand{\calD}{\mathcal{D}}
\newcommand{\calE}{\mathcal{E}}
\newcommand{\calF}{\mathcal{F}}
\newcommand{\calG}{\mathcal{G}}
\newcommand{\calR}{\mathcal{R}}

\newcommand{\Ebat}{\calE_\text{bat}}
\newcommand{\Ecap}{\calE_\text{cap}}

\newcommand{\DEbat}{\Delta\Ebat}
\newcommand{\DEcap}{\Delta\Ecap}

\newcommand{\Fe}{\calF_\text{e}}

\newcommand{\Rsigma}{\calR_\sigma}
\newcommand{\Rtheta}{\calR_\theta}

\newcommand{\tC}{\widetilde{C}}

\newcommand{\tF}{\widetilde{\calF}}
\newcommand{\tG}{\widetilde{\calG}}
\newcommand{\tR}{\widetilde{\calR}}

\newcommand{\tFe}{\widetilde{\calF}_\text{e}}
\newcommand{\tFr}{\widetilde{\calF}_\text{r}}

\newcommand{\tRtheta}{\widetilde{\calR}_\theta}
\newcommand{\tRsigma}{\widetilde{\calR}_\sigma}

\newcommand{\eps}{\epsilon}
\newcommand{\vareps}{\varepsilon}

\newcommand{\epsa}{\eps_\text{a}}
\newcommand{\epsr}{\eps_\text{r}}
\newcommand{\epara}{\eps_{\scriptscriptstyle\parallel}}
\newcommand{\eperp}{\eps_{\scriptscriptstyle\perp}}

\newcommand{\mur}{\mu_\text{r}}

\newcommand{\bfI}{\mathbf{I}}

\newcommand{\alphatensor}{\mathlarger{\bfalpha}}
\newcommand{\epstensor}{\mathlarger{\bfeps}}

\newcommand{\Freed}{Fr\'{e}edericksz}

\newcommand{\We}{W_\text{e}}
\newcommand{\WE}{W_\text{E}}

\newcommand{\Wcap}{W_{\text{cap}}}

\newcommand{\thz}{\theta_z}

\newcommand{\CC}{\cos^2\!\theta}
\renewcommand{\SS}{\sin^2\!\theta}

\newcommand{\KCCSS}{K_1\CC+K_3\SS}
\newcommand{\eCCSS}{\eperp\!\CC+\epara\SS}

\newcommand{\Knxnz}{K_1 n_{z,z}^2 + K_3 n_{x,z}^2}
\newcommand{\enxnz}{\eperp n_x^2+\epara n_z^2}

\newcommand{\ex}{\ehat_x}
\newcommand{\ey}{\ehat_y}
\newcommand{\ez}{\ehat_z}

\newcommand{\DU}{\Delta U}

\newcommand{\tauE}{\tau\!\raisebox{-.6ex}{\tiny{$E$}}}

\newcommand{\tausig}{\tau_\sigma}
\newcommand{\tauK}{\tau\!\raisebox{-.6ex}{\tiny{$K$}}}
\newcommand{\tauV}{\tau\!\raisebox{-.6ex}{\tiny{$V$}}}

\newcommand{\tauKbar}{\bar{\tau}\!\raisebox{-.6ex}{\tiny{$K$}}}
\newcommand{\tauVbar}{\bar{\tau}\!\raisebox{-.6ex}{\tiny{$V$}}}

\newcommand{\tauoff}{\tau_{\text{off}}}
\newcommand{\tauon}{\tau_{\text{on}}}

\newcommand{\Qf}{Q_\text{f}}
\newcommand{\DUf}{\DU_\text{f}}

\newcommand{\sigmaf}{\sigma_{\text{f}}}

\newcommand{\Kbar}{\bar{K}}

\newcommand{\tbar}{\bar{t}}
\newcommand{\zbar}{\bar{z}}

\newcommand{\sigbar}{\bar{\sigma}}

\newcommand{\td}{\text{d}}

\newcommand{\dt}{\td{t}}
\newcommand{\dtbar}{\td{\tbar}}
\newcommand{\dz}{\td{z}}
\newcommand{\dzbar}{\td{\zbar}}

\newcommand{\dQ}{\td{Q}}
\newcommand{\dS}{\td{S}}
\newcommand{\dV}{\td{V}}

\let\div\relax
\DeclareMathOperator{\div}{div}

\DeclareMathOperator{\curl}{curl}

\begin{document}

\title{A model liquid crystal cell in an RC circuit}

\author{Eugene C. Gartland, Jr.\thanks{Department of Mathematical
    Sciences, Kent State University, P.O.~Box 5190, Kent, OH 44242,
    USA (\myemail, \myurl)}}

\date{\today}

\maketitle

\begin{abstract}
  A nematic liquid-crystal cell subject to an electric field created
  by electrodes held at constant potential is modeled as a variable
  capacitor in an RC circuit.  The state of the system is
  characterized in terms of the director field in the cell and the
  charge on the electrodes.  A dynamical system is developed that
  couples director dynamics in the cell (with no fluid flow) and
  charge dynamics in the circuit.  The dynamical equations are derived
  from expressions for the total potential energy of the system and a
  dissipation involving a single rotational viscosity for the director
  plus Joule heating associated with current in the circuit.  An
  effort is made to quantify effects, in particular the widely
  differing time scales for the processes involved, and numerical
  illustrations are given.  The exercise illuminates aspects of the
  modeling of equilibrium states of such a system.
\end{abstract}


\section{Introduction}

We are motivated by a desire to understand better the dynamical
stability of equilibria of a nematic liquid crystal in an electric
field created by electrodes held at constant potential.  As modeled in
the Oseen-Frank macroscopic continuum theory, the free-energy
functional that governs the coupled equilibrium states of the liquid
crystal and the electric field can, in the simplest cases, be
expressed in the form
\begin{equation}\label{eqn:FnU}
  \calF[\nhat,U] = \int_\Omega \Bigl[ \We(\nhat,\nabla\nhat) -
  \frac12 \epstensor(\nhat) \nabla U \cdot \nabla U \Bigr] \, \dV .
\end{equation}
Here $\nhat$ is the director field (unit-length vector field), $U$ the
electric potential (related to the electric field via
$\bmE = - \nabla U$), $\Omega$ the domain of the liquid crystal cell,
$\We$ the density of distortional elastic energy, and
${\mathlarger\epstensor}$ the dielectric tensor.  These terms are
characterized more carefully in the next section.  The tensor
${\mathlarger\epstensor}$ is symmetric positive definite, and as a
result, the problem has an intrinsic ``minimax'' nature to it, with
locally stable equilibria locally minimizing with respect to $\nhat$
but maximizing with respect to $U$.  The assessment of local stability
is well understood from a variational point of view, studied in
\cite{gartland:21}.  However, the negative definiteness of $\calF$
with respect to $U$ causes confusion about how to assess stability
from a dynamical point of view.  The source of this confusion, as we
shall see, comes from the modeling assumptions that must be made in
order to put the free energy in the form above.  In particular, one
must assume either a static electric field or an electric field that
adjusts instantaneously to changes in the director field in order for
\eqref{eqn:FnU} to be valid.

The dynamics of such a system can be modeled at different levels of
fidelity.  A director field that is evolving in time will have
associated with it fluid flow in the cell, changes in the local
electric field (caused by changes in the dielectric tensor), and
changes in the capacitance of the cell (again caused by the changes in
the dielectric tensor), which will in turn cause changes in the charge
distributions on the electrodes.  Hydrodynamics is often of secondary
importance (and ignoring it greatly simplifies matters); so we shall
assume no fluid flow.  In many cases, the time scale for director
dynamics is orders of magnitude slower than the time scales for the
electric field and circuit dynamics; so a common modeling
approximation is to assume that both the electric field and the charge
distributions on the electrodes adjust instantaneously to any changes
in the director field.  This leaves a reduced free energy that is a
functional of $\nhat$ only, which makes the ``minimax problem'' go
away.  The equilibrium Euler-Lagrange equation for the reduced model
is nonlocal, however---the electric field at a point depends on the
director field everywhere---though the assessment of stability is
``normal,'' in that locally stable equilibrium director fields are
local minimizers of the reduced free energy.

There are, however, experiments involving ``fast switching''
(motivated by potential applications for light modulators and the
like) in which the time scale for director reorientation is comparable
to the time scale for circuit dynamics---see, for example,
\cite{geis:lyszczarz:osgood:kimball:10,
  gu:yin:shiyanovskii:lavrentovich:07,takanashi:maclennan:clark:98}.
There are, in fact, experiments in which the circuit dynamics are a
limiting factor \cite{baier-saip:bostanjoglo:eichler:macdonald:95,
  jang:clark:01}.  The time scale for the evolution of the electric
field, on the other hand, comes from the time-dependent Maxwell
equations and is invariably several orders of magnitude faster than
any other macroscopic time scale present.  We examine these issues
more carefully in \S\ref{sec:TimeScales}.

Here then we assume that the electric field adjusts instantaneously to
the director field, but we model faithfully the coupling between the
dynamics of the director field and those of the charge distribution.
The modeling highlights the role of the voltage source in giving rise
to the troubling minus sign in the free energy in \eqref{eqn:FnU}
above and the assumptions that must be made in order to put the free
energy in that form.  Also, the coupling between director dynamics and
circuit dynamics introduces an additional mechanism for energy
dissipation: Joule heating due to current in the circuit.

We choose a ``textbook'' model system for illustration (a
splay-\Freed\ cell) so that all formulas can be worked out explicitly
and are not too cumbersome.  Several of our results apply to more
general situations, and we try to indicate this at appropriate points.
The model problem, free energy, and equilibrium equations are
presented in \S\ref{sec:model}.  There, both the coupled formulation
(in terms of $\nhat$ and $U$) and the reduced formulation (in terms of
$\nhat$ only) are discussed.  The basic equations characterizing RC
circuits are reviewed in \S\ref{sec:RCcircuits}.  In
\S\ref{sec:joint}, the linkage is made between the state of the
director field in the cell and the state of the electric circuit, with
a combined potential energy and dissipation function and resulting
coupled dynamical system.  A simple numerical illustration of the
coupled dynamics is given in \S\ref{sec:numerics}, and what
conclusions can be drawn are discussed in \S\ref{sec:conclusions}.

\section{Model and equilibrium equations}\label{sec:model}

\begin{figure}
  \centering
  \subfloat[]{\label{fig:geom} 
    \includegraphics[width=.49\linewidth]{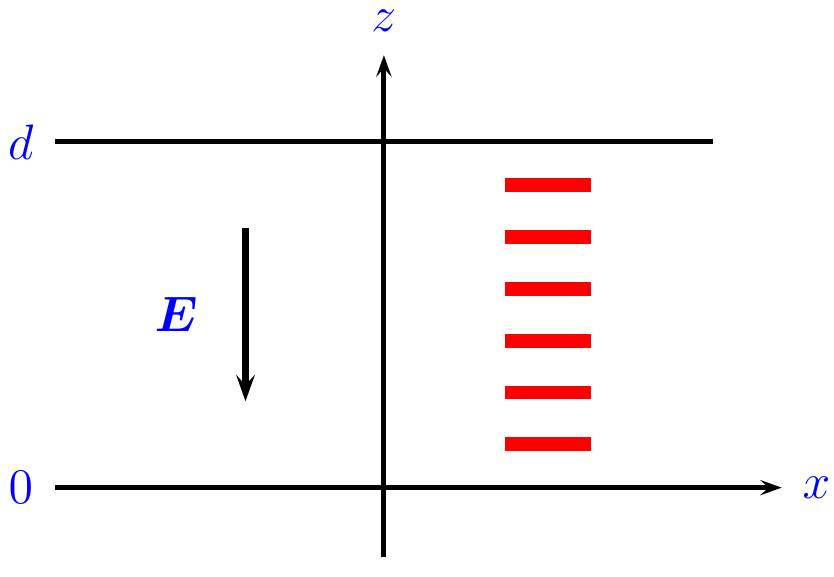}}
  \subfloat[]{\label{fig:RCcircuit}
    \raisebox{3.25ex}{\includegraphics[width=.49\linewidth]{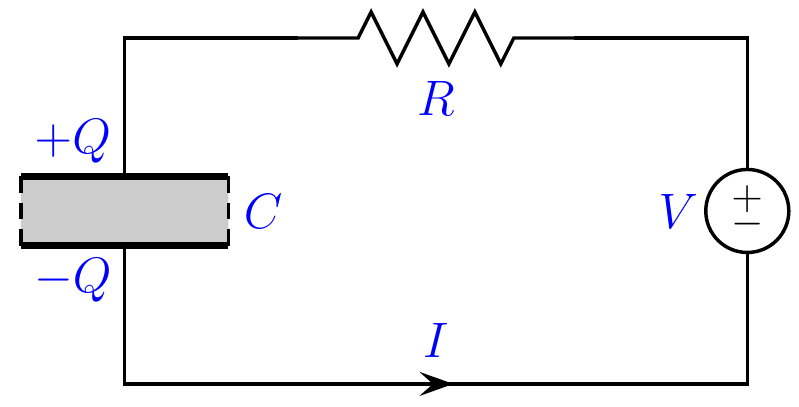}}}
  \caption{Model problem.  Figure\,\ref{fig:geom}: electric-field
    splay-\Freed-geometry cell (geometry, coordinate system, ground
    state) as a capacitor in an RC circuit,
    figure\,\ref{fig:RCcircuit}.  The liquid crystal film is confined
    to $0<z<d$.  Strong anchoring is assumed on $z=0$ and $z=d$, and
    the liquid-crystal director is assumed to remain in
    $\operatorname{span}\{\ex,\ez\}$.  RC circuit parameters:
    resistance $R$, capacitance $C$, voltage $V$, charge $Q$ (upper
    electrode $+$, lower electrode $-$), current $I$ (positive
    direction indicated).}
\end{figure}

For a concrete realization of the ideas explored here, we consider a
nematic liquid crystal cell in the splay-\Freed\ geometry (as depicted
in figure\,\ref{fig:geom}) subject to an electric
field created by electrodes held at constant potential by a voltage
source.  In devices, the voltage source is usually just a battery,
while in experiments, the electromotive force could come from a
variable power supply.  The system is modeled using the Oseen-Frank
macroscopic continuum theory \cite[Ch.\,3]{degennes:prost:93},
\cite[Ch.\,2]{stewart:04}, \cite[Ch.\,3]{virga:94}.  The lateral
dimensions of the cell are assumed to be much larger than the cell
gap, enabling us to treat all fields as uniform in $x$ and $y$ (the
coordinates in the lateral directions), with spatial dependence only
on the $z$ coordinate (the coordinate across the cell gap)---thus we
ignore the influence of fringe fields and effects near the edges of
the cell.  We assume strong anchoring conditions and further assume
that the director $\nhat$ remains in the $x$-$z$ tilt plane:
\begin{equation*}
  \nhat = n_x \ex + n_z \ez =
  \cos \theta \, \ex + \sin \theta \, \ez , \quad
  \theta = \theta(z) .
\end{equation*}

The free-energy functional will contain contributions from
distortional elasticity plus terms associated with the electric
field.  The elastic part has the form
\begin{equation}\label{eqn:Fe}
  \Fe[\nhat] = \int_\Omega \We(\nhat,\nabla\nhat) \, \dV =
  A \int_0^d \We(\nhat,\partial_z\nhat) \, \dz ,
\end{equation}
where $\Omega$ is the domain of the cell, $A$ the $x$-$y$
cross-section area, $d$ the cell gap, and
\begin{equation}\label{eqn:We}
  \begin{aligned}
    2 \We &= K_1 (\div\nhat)^2 + K_2 (\nhat\cdot\curl\nhat)^2 +
             K_3 |\nhat\times\curl\nhat|^2 \\
          &= \Knxnz \\
          &= \bigl( \KCCSS \bigr) \thz^2 .
  \end{aligned}
\end{equation}
See \cite[\S3.1.2]{degennes:prost:93}, \cite[\S2.2.1]{stewart:04},
\cite[\S3.2]{virga:94}.  Here we denote $n_{x,z}=\td n_x/ \dz$,
$n_{z,z}=\td n_z/\dz$, and $\thz=\td\theta/\dz$.  Thus the
distortional free energy per unit area, $\tFe := \Fe / A$, as a
function of the tilt angle $\theta$ is given by
\begin{equation*}
  \tFe[\theta] = \frac12 \int_0^d \bigl(
  \KCCSS \bigr) \thz^2 \, \dz .
\end{equation*}

The appropriate contribution to the free-energy density associated
with a static electric field arising from electrodes held at constant
potential is
\begin{equation*}
  \WE = - \frac12 \bmD \cdot \bmE ,
\end{equation*}
where $\bmD$ is the displacement field and $\bmE$ the electric field.
This is discussed in a general context in \cite[\S4.7]{jackson:75} and
\cite[\S5, \S10]{landau:lifshitz:pitaevskii:93}; it is discussed in
the particular context of liquid crystals in \cite[\S3.6.2,
\S3.6.3]{barbero:evangelista:01}, \cite[\S10.1]{collings:hird:97}, and
\cite[\S7.1]{yang:wu:15}.  Another perspective on this expression will
be developed in \S\ref{sec:system-potential-energy}.  In a system such
as ours (a transversely isotropic medium in the linear regime), it is
usually assumed that
\begin{equation}\label{eqn:DepsE}
  \bmD = \epstensor(\nhat) \bmE , \quad
  \epstensor = \eps_0 \bigl[ \eperp \bfI +
  \epsa ( \nhat \otimes \nhat ) \bigr] , \quad
  \epsa := \epara - \eperp ,
\end{equation}
with ${\mathlarger\epstensor}$ the dielectric tensor and $\eperp$ and
$\epara$ the relative dielectric permittivities perpendicular to
$\nhat$ and parallel to $\nhat$, giving
\begin{equation*}
  \WE = - \frac12 \eps_0 \bigl[ \eperp E^2 +
  \epsa ( \bmE\cdot\nhat )^2 \bigr] , \quad E = | \bmE | .
\end{equation*}
The macroscopic modeling of electric fields in liquid crystals is
discussed in \cite[\S3.3.1]{degennes:prost:93},
\cite[\S2.3.1]{stewart:04}, and \cite[\S4.1]{virga:94}.  We note that
the electric field is in general nonhomogeneous \cite{gartland:21} and
that the linear relation between static $\bmD$ and $\bmE$ fields can
be nonlocal in space, though such spatial dispersion is generally
viewed as negligible in macroscopic models of pure dielectrics---see
\cite[\S I.4]{jackson:75} or \cite[\S103]{landau:lifshitz:pitaevskii:93}.

The relevant Maxwell equations for electrostatics (assuming no
distribution of free charge in $\Omega$) are
\begin{equation*}
  \curl \bmE = \bfzero , \quad \div \bmD = 0 .
\end{equation*}
Tangential components of $\bmE$ are continuous across a material
interface, while the normal component of $\bmD$ suffers a jump equal
to the surface charge density $\sigmaf$ on the interface:
\begin{equation*}
  \llbracket E_{\text{t}} \rrbracket = 0 , \quad
  \llbracket D_\nu \rrbracket = \sigmaf .
\end{equation*}
Both $\bmD$ and $\bmE$ vanish inside electrodes.  The basic
electrostatics that we require can be found in any of \cite[Ch.\,I,
Ch.\,1]{jackson:75}, \cite[Ch.\,I,
Ch.\,II]{landau:lifshitz:pitaevskii:93},
\cite[Chs.\,2--6]{reitz:milford:67}, or \cite[Ch.\,III]{stratton:41}.
Using $\curl\bmE=\bfzero$ and the interface conditions, we conclude
that in our system
\begin{equation}\label{eqn:EUzDzsigma}
  \bmE = - U_z \ez , \quad U_z = \td U / \dz , \quad
  D_z(0+) = D_z(d-) = - \sigma ,
\end{equation}
where $U=U(z)$ is the electric potential and $\sigma$ is the (uniform)
surface charge density on the upper electrode ($-\sigma$ on the lower
electrode).  Given the polarity indicated in our circuit diagram in
figure\,\ref{fig:RCcircuit}, we have $\sigma > 0$.  Thus $\WE$
simplifies to
\begin{equation*}
  \WE = - \frac12 \eps_0 \bigl( \enxnz \bigr) U_z^2 .
\end{equation*}
For equilibrium states of our model problem, then, the total free
energy (per unit area) expressed as a functional of $\theta$ and $U$
is given by
\begin{equation}\label{eqn:Fcoupled}
  \tF[\theta,U] = \frac12 \int_0^d \bigl[ \bigl(\KCCSS\bigr) \thz^2 -
  \eps_0 \bigl(\eCCSS\bigr) U_z^2 \bigr] \, \dz .
\end{equation}

\subsection{Coupled system}

When viewed as a coupled system in this way, the equilibrium
Euler-Lagrange equations that follow from $\delta_\theta\tF=0$ and
$\delta_U\tF=0$ are given by
\begin{subequations}\label{eqn:coupled-system}
  \begin{gather}
    \frac{\td}{\dz} \bigl[ \bigl(\KCCSS\bigr) \thz \bigr] =
    \sin\theta \cos\theta \bigl[ (K_3-K_1) \thz^2 -
    \eps_0 \epsa U_z^2 \bigr] ,
    \label{eqn:thetaODE} \\
    \frac{\td}{\dz} \bigl[ \bigl(\eCCSS\bigr) U_z \bigr] = 0 ,
    \label{eqn:UODE}
  \end{gather}
  with boundary conditions
  \begin{equation}
    \theta(0) = \theta(d) = 0 , \quad
    U(0) = - \DU / 2 , ~ U(d) = \DU / 2 .
  \end{equation}
\end{subequations}
Here $\DU$ is the difference between the electric potential on the
upper electrode and that on the lower electrode, and the director
field and electric field in the cell depend only on this
difference---an arbitrary constant added to $U$ has no effect on
solutions of \eqref{eqn:coupled-system} (one could just as well impose
boundary conditions $U(0)=0$, $U(d)=\DU$).  Equation \eqref{eqn:UODE}
corresponds to $\div\bmD=0$ for our model cell and emerges in a
natural way as a condition of stationarity of $\tF[\theta,U]$.

The characterization of the stability properties of solutions of the
coupled system \eqref{eqn:coupled-system} is a little nonstandard
because of the ``minimax'' nature of the free energy $\tF$ in
\eqref{eqn:Fcoupled}.  Locally stable solutions are locally minimizing
with respect to $\theta$ but maximizing with respect to $U$; while
globally stable solutions are globally minimizing with respect to
$\theta$, maximizing with respect to $U$.  Globally stable solutions
can also be characterized as equilibrium solutions of least free
energy.  These topics are taken up in \cite{gartland:21}.  From the
point of view of \emph{dynamical} stability, the picture is somewhat
confusing, and a goal of this note is to try to understand this
better.

\subsection{Reduced system}

Instead of viewing the system as a coupled system with two state
variables, one can eliminate $U$ and model the system in terms of
$\theta$ only.  An integration of \eqref{eqn:UODE} gives
\begin{equation}\label{eqn:Uz}
  U_z = \DU \biggl[ \int_0^d \frac{\dz}{\eCCSS} \biggr]^{-1}
  \!\! \frac1{\eCCSS} \, ,
\end{equation}
which when substituted into \eqref{eqn:thetaODE} produces
\begin{multline}\label{eqn:thetaODEnonlocal}
  \frac{\td}{\dz} \bigl[ \bigl(\KCCSS\bigr) \thz \bigr] =
  \sin\theta \cos\theta \biggl\{ (K_3-K_1) \thz^2 \\
  {} - \eps_0 \epsa (\DU)^2 \biggl[ \int_0^d \frac{\dz}{\eCCSS}
  \biggr]^{-2} \!\! \frac1{(\eCCSS)^2} \biggr\} .
\end{multline}
This same equation can be obtained from the first variation of the
reduced free energy $\tFr$ that results when the expression for $U_z$
above is substituted into \eqref{eqn:Fcoupled}:
\begin{equation}\label{eqn:tFr}
  \tFr[\theta] = \frac12 \int_0^d \bigl(\KCCSS\bigr) \thz^2 \, \dz -
  \frac12 \eps_0 (\DU)^2 \biggl[ \int_0^d \frac{\dz}{\eCCSS}
  \biggr]^{-1} \!\! .
\end{equation}
If the voltage of the battery is $V$, then in equilibrium
$\DU=V$, and the above expression agrees with
\cite[(3.221)]{stewart:04}.  When the system is \emph{not} in
equilibrium, however, $\DU$ need not be equal to $V$, and this
will be an issue in what follows.

The formulation in terms of the reduced free energy is natural and has
been widely used.  It was used by Deuling in \cite{deuling:72}, which
is recounted in \cite[\S3.5]{stewart:04}.  It was used by Hardt,
Kinderlehrer, and Lin in their analytical paper
\cite{hardt:kinderlehrer:lin:86}, and it was also employed by the
author in \cite{gartland:21}.  The formulation has certain advantages
in terms of stability assessment, in that locally stable states are
local minimizers of $\tFr$ with respect to $\theta$, and globally
stable states are global minimizers (the second variation
$\delta^2\!\tFr$ being positive definite in both cases).  Using $\tFr$
is equivalent to minimizing the free energy $\tF$ in
\eqref{eqn:Fcoupled} subject to the ODE constraint \eqref{eqn:UODE}.
The approach amounts to viewing the electric field as slaved to the
director field.  A disadvantage of the reduced-free-energy formulation
is that the equilibrium Euler-Lagrange equation
\eqref{eqn:thetaODEnonlocal} is \emph{nonlocal}.

When the system is out of equilibrium and the director field is
evolving in time ($\nhat=\nhat(z,t)$, $\theta=\theta(z,t)$), then the
capacitance of the liquid-crystal cell will be changing with time as
well.  In this case, the battery will need to move charge on or off
the electrodes in order to re-establish the equilibrium potential
difference $\DU = V$.  The associated currents in the electric circuit
will suffer some energy loss, due to Joule heating, and we need a way
to combine these effects with the dynamics and viscous dissipation in
the liquid-crystal cell.  We begin by reviewing the charge dynamics of
a standard RC circuit.

\section{RC circuits}\label{sec:RCcircuits}

The typical kind of experimental setup that we envision can be viewed
as an RC circuit with the liquid-crystal cell forming a capacitor
containing a complex time-varying dielectric, as depicted in
figure\,\ref{fig:RCcircuit}.  The resistance $R$ could come from the
presence of an actual resistor, or it could just be thought of as a
surrogate to account for the total resistance of all the elements in
the circuit (wires, connectors, conduction layers in the cell, etc.).
The equation governing charge dynamics in such a circuit follows from
the Kirchhoff Voltage Law and the formulas for the potential drops
across a resistor and a capacitor \cite[\S6.6,
\S7.8]{reitz:milford:67}:
\begin{equation*}
  \DU_{\text{res}} = I R , ~ \DU_{\text{cap}} = Q / C ~ \Rightarrow ~
  I R + Q / C = V .
\end{equation*}
Here $Q(t)$ is the instantaneous total charge on the upper (positive)
electrode, and $I = \dQ/\dt$ is the current.  Appending an initial
condition leads to the IVP
\begin{equation}\label{eqn:QODEIVP}
  R \frac{\dQ}{\dt} + \frac1{C} Q = V , ~~ Q(0) = Q_0 ,
\end{equation}
the solution of which can be written
\begin{equation}\label{eqn:Qt}
  Q(t) = Q_\infty + ( Q_0 - Q_\infty ) \exp(-t/\tau) , \quad
  Q_\infty := C V , ~~ \tau := RC .
\end{equation}
Thus the steady state of the circuit corresponds to
\begin{equation*}
  Q = C V , ~~ \DU = V ,
\end{equation*}
and the characteristic time scale for the dynamics is $\tau =
RC$---from here on, we simply denote $\DU_{\text{cap}} = \DU$ (as we
have used in the previous section).

The state of the circuit could just as well be characterized in terms
of $\DU$ (instead of $Q$), in which case \eqref{eqn:QODEIVP} would
take the form
\begin{equation*}
  R \frac{\td}{\dt} ( C \DU ) + \DU = V , ~~ \DU(0) = \DU_0 .
\end{equation*}
As we shall see in what follows, $C$ depends on $\nhat$ (which will be
changing in time, leading to $C=C(t)$); so modeling the circuit in
terms of $Q$ proves to be more convenient (at least for our model
problem).

The potential energies of the capacitor and battery, relative to a
value of zero at $Q=0$, are given by
\begin{equation*}
  \Ecap = \frac12 Q \DU , ~~ \Ebat = - Q V .
\end{equation*}
The relation $Q=C\DU$ makes it possible to write these expressions in
various equivalent forms, e.g., $\Ecap=Q^2/2C$ or $\Ecap=C(\DU)^2/2$.
These potential energies correspond to the work done in a reversible
process building up the charge on the capacitor electrodes in an
incremental way.  As the charge on the capacitor is being built up,
the electric potential is changing there as well (according to the
relation $Q = C \DU$), leading to the factor of $1/2$: the increment
of work done in moving an increment of charge from a location of
potential zero to a location of potential $\DU$ is $\delta W = \delta
Q \DU$, giving
\begin{equation*}
  \Wcap =
  \int_0^{\Qf} \!\! \DU \, \dQ =
  \int_0^{\Qf} \! \frac{Q}{C} \, \dQ =
  \frac1{C} \int_0^{\Qf} \!\! Q \, \dQ =
  \frac1{2C} \Qf^2 =
  \frac12 \Qf \DUf ,
\end{equation*}
where $\Qf$ and $\DUf$ are the final (fully charged) values of $Q$ and
$\DU$.  The battery, on the other hand, always maintains a constant
potential of $V$; so
\begin{equation*}
  W_{\text{bat}} = \int_{\Qf}^0 \! V \, \dQ =
  V \! \int_{\Qf}^0 \! \dQ = - \Qf V .
\end{equation*}
A discussion of these textbook formulas can be found in
\cite[\S6.6]{reitz:milford:67}.  By convention, the accounting is done
in terms of just the positive electrode, but it takes into account the
contribution of the negative electrode---in actuality, one has charge
$Q$ on the positive electrode at potential $\DU/2$ and charge $-Q$ on
the negative electrode at potential $-\DU/2$.  We conclude that the
total potential energy at any instant, expressed in terms of $Q$, is
given by
\begin{equation*}
  \calE := \Ecap + \Ebat = \frac1{2C} Q^2 - Q V .
\end{equation*}
The discussion above assumes a \emph{constant} capacitance $C$.  We
will revisit this calculation later with $C = C(Q)$.

We note that whether the system is in equilibrium or not, we always
have equal and opposite total excess charge on the upper and lower
electrode surfaces.  This relates to conservation of charge and is a
consequence of Gauss's Law.  In the full time-dependent Maxwell
equations (as well as in Maxwell electrostatics), we always have
$\div\bmD=0$ in the absence of any free-charge distribution in
$\Omega$ (which we have assumed to be the case throughout).  From this
follows
\begin{equation*}
  0 = \int_\Omega \div \bmD \, \dV =
  \int_{\partial\Omega} \! \bmD \cdot \nuhat \, \dS =
  - \int_{\partial\Omega} \! \sigmaf \, \dS .
\end{equation*}
In our system, $\sigmaf$ is supported on the top and bottom electrode
surfaces; so the total charge on the top and bottom must sum to zero
($Q$ on the top and $-Q$ on the bottom, in our notation).  We also
note that we use interchangeably the terms ``potential energy,''
``electrostatic energy,'' and ``free energy''---all refer to the
energy associated with reversible work processes.

Current flowing through a resistor leads to energy dissipation (by
Joule heating) at a rate $RI^2$---see
\cite[(21.6)]{landau:lifshitz:pitaevskii:93} or
\cite[\S1.8, (4)]{stratton:41}.  The Rayleigh dissipation
function associated with this is
\begin{equation}\label{eqn:calRcircuit}
  \calR = \frac12 R I^2 = \frac12 R \Bigl( \frac{\dQ}{\dt} \Bigr)^{\!2} .
\end{equation}
From a variational point of view, then, the circuit dynamics can be
obtained from a dissipation principle:
\begin{equation*}
  \frac{\partial\calE}{\partial Q} +
  \frac{\partial\calR}{\partial\dot{Q}} = 0 ~ \Rightarrow ~
  \frac1{C} Q - V + R \frac{\dQ}{\dt} = 0 ,
\end{equation*}
in agreement with \eqref{eqn:QODEIVP}.  Here we follow the formalism
of Lagrangian mechanics with frictional forces and potential energy
only (no kinetic energy), denoted in that setting $L=T-V=-V$ (with
$V=\calE$ here)---see, for example, \cite[\S2.2.1]{sonnet:virga:12},
where a historical perspective and classical references can be found.

One can try to put some of this in context.  The total energy
dissipated in the dynamical process is given by
\begin{equation*}
  \calD \! = \int_0^\infty \!\! R I^2 \, \dt = \frac1{2C} ( Q_\infty - Q_0 )^2 ,
\end{equation*}
using \eqref{eqn:Qt}, compared to the potential-energy changes of the
capacitor and battery:
\begin{equation*}
  \DEcap = \frac1{2C} \bigl( Q_\infty^2 - Q_0^2 \bigr) ,
  \quad \DEbat = - ( Q_\infty - Q_0 ) V =
  \frac1{C} ( Q_0 - Q_\infty ) Q_\infty .
\end{equation*}
Thus we always have
\begin{equation*}
  \DEcap + \DEbat + \calD = 0 ,
\end{equation*}
as we must.  In the special case of charging the capacitor from
\emph{zero} ($Q_0=0$), we have
\begin{equation*}
  \calD = \DEcap = \frac12 C V^2 , \quad \DEbat = - C V^2 .
\end{equation*}
In this case, half the work done by the battery goes into the final
electrostatic energy of the capacitor, while the other half is lost to
dissipation.  Similar phenomena (in which half of the work is lost to
dissipation) occur in a number of other settings, including simple
spring-mass-damper systems and linear elasticity
\cite{fosdick:truskinovsky:03}.

At the other extreme, if the initial charge on the
capacitor were just slightly out of equilibrium,
$Q_0 = ( 1 + \vareps ) Q_\infty $, say, then one would obtain
\begin{equation*}
  \DEcap = - \vareps ( 1 + \vareps/2 ) C V^2 , \quad
  \DEbat = \vareps C V^2 , \quad
  \calD = \vareps^2 C V^2 / 2 .
\end{equation*}
In this case, both $\DEcap$ and $\DEbat$ are $O(\vareps)$, but
$\calD = O(\vareps^2)$---and the dissipation would essentially be
negligible when $|\vareps|\ll1$.  While we always have
\begin{equation*}
  0 \le \calD \le | \DEcap | ,
\end{equation*}
we only have $\calD = | \DEcap |$ in the case $Q_0 = 0$ or in the
trivial cases $Q_0=Q_\infty$ (no dynamics) or $Q_\infty=0$ (no
battery).

\section{Total potential energy and dynamical system}

\label{sec:joint}

There is a mutual influence between the director field in the cell and
the state of the electric circuit: the director field determines the
capacitance of the cell (which affects the circuit), while the state
of the circuit (whether characterized by $Q$ or by $\DU$) affects the
electric field in the cell (and hence the director field).  In order
to put together a coupled set of equilibrium equations or dynamical
equations, we require expressions for the total potential energy and
for the dissipation of the full system.  These, in turn, require
certain ``building blocks,'' which we now derive.

\subsection{Capacitance of the cell}

Due to the simplicity of our model problem (fields depending on only
one space variable, stratified nature of the dielectric), it is
possible to express the capacitance of the cell in analytical form:
\begin{equation}\label{eqn:Cn}
  C[\nhat] = \eps_0 A \biggl[
  \int_0^d \!\! \frac{\dz}{\enxnz} \biggr]^{-1} \!\! .
\end{equation}
This can be derived in various ways.  Using the basic relation for a
parallel-plate capacitor $Q=C\DU$ (with $Q$ the total charge on the
positive electrode, $C$ the capacitance, and $\DU$ the potential
difference between the electrodes, as before) and the relations in
\eqref{eqn:DepsE} and \eqref{eqn:EUzDzsigma}, we have
\begin{equation*}
  D_z = \eps_{zz} E_z = - \sigma ~ \Rightarrow ~
  U_z = - E_z = \frac{\sigma}{\,\eps_{zz}} ~ \Rightarrow ~
  \DU = \! \int_0^d \! U_z \, \dz =
  \sigma \! \int_0^d \! \frac{\dz}{\,\eps_{zz}} \, ,
\end{equation*}
which gives
\begin{equation*}
  C = \frac{Q}{\DU} =
  \frac{\sigma A}{\sigma \! \int_0^d (1/\eps_{zz}) \, \dz} =
  A \biggl[ \int_0^d \! \frac{\dz}{\,\eps_{zz}} \biggr]^{-1} \!\!\! =
  \eps_0 A \biggl[ \int_0^d \!\! \frac{\dz}{\enxnz} \biggr]^{-1} \!\! .
\end{equation*}
In this situation, $C$ is simply an integral functional of the
director field.  One also sees such expressions derived by
approximating the stratified dielectric as a collection of thin
capacitive elements in series \cite[\S5.4.3.1]{dunmur:toriyama:99}.
Note that if the material were isotropic with relative permittivity
$\epsr$ (i.e., $\eperp=\epara=\epsr$), then \eqref{eqn:Cn} would
simplify to $C=\eps A/d$ with $\eps=\eps_0\epsr$, which is the
textbook expression for the capacitance of a parallel-plate capacitor
\cite[\S6.6]{reitz:milford:67}.

In more general circumstances (such as a director field that is a
function of more than one space variable), it is not possible to
derive a formula such as the above: there is no explicit analytical
expression for the capacitance of a cell with a two- or
three-dimensional inhomogeneity of the dielectric.  In such cases, one
can characterize the capacitance in terms of the director field
$\nhat$ in $\Omega$ given either $Q$ or $\DU$, but to determine $C$,
one must solve an auxiliary problem from an appropriate formulation of
Maxwell electrostatics, then determine $\DU$ from this solution (if
$Q$ was given) or $Q$ (if $\DU$ was given), and finally compute the
ratio $C=Q/\DU$.

\subsection{Potential energy of the cell}

\label{sec:potential-energy-cell}

The textbook calculation of the electrostatic energy of a capacitor
was recounted in \S\ref{sec:RCcircuits}.  When combined with the
relation $Q=C\DU$, it yielded several equivalent ways of writing this
potential energy:
\begin{equation}\label{eqn:Ecap}
  \Ecap = \frac12 Q \DU = \frac1{2C} Q^2 = \frac12 C (\DU)^2 .
\end{equation}
The calculation of $\Ecap$ in \S\ref{sec:RCcircuits} depended on the
assumption that the capacitance of the cell was \emph{constant}.  In
our system, however, the capacitance changes with changes in the
director field in the cell, and we must modify this calculation
accordingly.

As before, we start from the fact that the increment of work
$\delta W$ done in moving an increment of charge $\delta Q$ from one
location to another location of potential difference $\DU$ is
$\delta W = \delta Q \DU$, giving (as before)
\begin{equation*}
  \Wcap = \int_0^{\Qf} \!\! \DU \, \dQ =
  \int_0^{\Qf} \! \frac{Q}{C} \, \dQ .
\end{equation*}
Now, however, $C$ depends on $Q$ (in a way that could be complicated
and not easy to express)---as charge is added to the upper electrode
(depleted from the lower electrode), the electric field in the cell
will eventually become strong enough to distort the director field
(and change the capacitance).  An integration by parts provides a
simple way to assess the new situation:
\begin{multline*}
  \Wcap = \int_0^{\Qf} \!\! \frac{Q}{C(Q)} \, \dQ =
  \int_0^{\Qf} \!\! \frac1{C(Q)} \, \td \Bigl( \frac{Q^2}2 \Bigr) \\ =
  \frac12 \frac{\Qf^2}{C(\Qf)} + \frac12 \int_0^{\Qf} \!\!
  \frac{Q^2}{C(Q)^2} \, \frac{\td C}{\dQ} \, \dQ =
  \frac12 \frac{\Qf^2}{C(\Qf)} + \frac12 \int_{C(0)}^{C(\Qf)} \!\!
  ( \DU )^2 \, \td C .
\end{multline*}
The first term on the right-hand side above is one form of the
electrostatic energy $\Ecap$ of a capacitor with a constant
capacitance $C=C(\Qf)$, as in \eqref{eqn:Ecap}.  The second term above
can be interpreted as follows.

The increment of work $\delta W$ done in changing the capacitance by
an incremental amount $\delta C$ is $\delta W=\frac12\delta C(\DU)^2$,
which can be seen from the alternate form of the potential energy of
the capacitor $\Ecap=\frac12C(\DU)^2$.  Thus the integral term on the
right-hand side above is the reversible work done in changing the
capacitance of the cell from its initial value at $Q=0$ to its final
value at $Q=\Qf$.  In our system, the only way the capacitance can be
changed is by distorting the director field in $\Omega$, and the work
function for that process is the distortional elastic energy $\Fe$.
Thus the integral in question gives the change in distortional elastic
energy from its value at $Q=0$ to its value at $Q=\Qf$:
\begin{equation*}
  \frac12 \int_{C(0)}^{C(\Qf)} \!\! ( \DU )^2 \, \td C =
  \Fe\bigr|_{Q=\Qf} - \Fe\bigr|_{Q=0} = \Fe\bigr|_{Q=\Qf} ,
\end{equation*}
since $\Fe|_{Q=0}=0$ in our system.  Under more general circumstances,
one could have $\Fe|_{Q=0}\not=0$, but that term would just add a
constant to the potential energy and could be ignored.

We see that the potential energy of the liquid-crystal cell in our
system is given by
\begin{equation*}
  \Wcap = \Ecap + \Fe[\nhat] , \quad
  \Ecap = \frac12 Q \DU =
  \frac12 C[\nhat]^{-1} Q^2 =
  \frac12 C[\nhat] ( \DU )^2 ,
\end{equation*}
where $\nhat$, $Q$, and $\DU$ are the instantaneous values of these
state variables.  This expression captures, in a clean and decoupled
way, both the work done in moving charge and that done by inducing
change in the capacitance (by distorting the director field).  We note
that we continue to use the notation $\Ecap$ for any of the equivalent
formulas for the electrostatic energy of a capacitor of constant
capacitance \eqref{eqn:Ecap}, with the understanding that the value of
$C$ is always taken to be that associated with the current state of
the director field: $C=C[\nhat]$.

\subsection{Potential energy of the system and coupled equilibrium equations}

\label{sec:system-potential-energy}

The total potential energy of the system $\calG$ is thus given by
\begin{equation}\label{eqn:calG}
  \calG = \Fe + \Ecap + \Ebat = \Fe + \frac12 Q \DU - Q V .
\end{equation}
The first two terms represent the potential energy of the
liquid-crystal cell, as derived in \S\ref{sec:potential-energy-cell},
while the third term is the potential energy associated with the
battery.  The expression for $\calG$ can be related to more familiar
forms by noting that for static electric fields, $\Ecap$ can be
expressed in terms of field intensities inside the cell:
\begin{equation}\label{eqn:QDUvsDE}
  \frac12 Q \DU = \frac12 \int_\Omega (\bmD\cdot\bmE) \,  \dV ,
\end{equation}
where $\bmD$ and $\bmE$ are the fields associated with the
electrostatic problem $\div\bmD=0$ in $\Omega$ with a potential
difference $\DU$ between the electrodes.  This can be established as
follows.  Using the relations $\bmD\cdot\nuhat = -\sigmaf$ (with
$\nuhat$ the outward normal from $\Omega$), $\div\bmD=0$, and
$\bmE = -\nabla U$, we obtain the following relation for a general
surface charge density $\sigmaf$ and electric potential $U$:
\begin{multline*}
  \int_{\partial\Omega} \! \sigmaf \, U \, \dS =
  - \int_{\partial\Omega} ( U \bmD ) \cdot \nuhat \, \dS =
  - \int_\Omega \div ( U \bmD ) \, \dV \\ =
  - \int_\Omega ( U \div \bmD + \nabla U \cdot \bmD ) \, \dV =
  \int_\Omega ( \bmD \cdot \bmE ) \, \dV .
\end{multline*}
See, for example, \cite[\S6.3]{reitz:milford:67} or
\cite[\S2.8]{stratton:41}.  In our system, the surface charge density
is supported on the upper and lower boundary electrodes, and the
electric potential $U$ is constant on each electrode ($-\DU/2$ on the
lower, $\DU/2$ on the upper), giving
\begin{multline*}
  \int_{\partial\Omega} \! \sigmaf \, U \, \dS =
  \int_{\Gamma_1} \! \sigmaf \, U \, \dS +
  \int_{\Gamma_2} \! \sigmaf \, U \, \dS =
  - \frac{\DU}2 \int_{\Gamma_1} \! \sigmaf \, \dS
  + \frac{\DU}2 \int_{\Gamma_2} \! \sigmaf \, \dS \\ =
  - \frac{\DU}2 (-Q) + \frac{\DU}2 Q = Q \DU .
\end{multline*}
Here $\Gamma_1$ and $\Gamma_2$ are the lower and upper boundary
electrode interfaces.  Combining these two calculations establishes
the validity of \eqref{eqn:QDUvsDE}.  Note that this argument does not
require $\sigmaf$ to be constant on $\Gamma_1$ or $\Gamma_2$.

Concerning these equivalent formulas for electrostatic energy
\begin{equation*}
  \frac12 \int_{\partial\Omega} \! \sigmaf \, U \, \dS =
  \frac12 \int_\Omega ( \bmD \cdot \bmE ) \, \dV ,
\end{equation*}
the left-hand side is in fact the more primitive expression (how
electrostatic energy associated with surface charge distributions is
often first derived in electromagnetics textbooks)---see, for example,
\cite[\S6.2]{reitz:milford:67} or \cite[\S2.7]{stratton:41}.  We note
that when there is no current flowing in the circuit, then $\DU = V$,
and the combination $\Ecap+\Ebat$ satisfies
\begin{equation}\label{eqn:calEequilib}
  \Ecap + \Ebat = \frac12 Q V - Q V = - \frac12 Q V =
  - \frac12 \int_\Omega ( \bmD \cdot \bmE ) \, \dV ,
\end{equation}
which is the appropriate contribution to the free energy to be used in
modeling liquid crystal equilibrium states in a setting such as ours.
We emphasize that in order for \eqref{eqn:calEequilib} to be valid,
one requires equilibrium conditions in the electrical circuit ($I=0$,
$\DU=V$) and also equilibrium of the electric field in the cell
($\curl\bmE=\bfzero$, $\bmE=-\nabla U$)---with time-varying electric
fields, the time-dependent Maxwell equations have
$\curl\bmE\not=\bfzero$, in general, and $\bmE$ cannot be expressed as
the gradient of a scalar potential.

The expression for $\calG$ in \eqref{eqn:calG} can be written in
different forms, depending upon the choice of state variable for the
circuit:
\begin{subequations}
\begin{equation}\label{eqn:GnDU}
  \calG[\nhat,\DU] = \Fe[\nhat] +
  C[\nhat] \Bigl[ \frac12 (\DU)^2 - V \DU \Bigr]
\end{equation}
or
\begin{equation}\label{eqn:GnQ}
  \calG[\nhat,Q] = \Fe[\nhat] + \frac12 C[\nhat]^{-1} Q^2 - V Q .
\end{equation}
\end{subequations}
Here $\nhat$ is the current state of the director field in $\Omega$,
$\Fe[\nhat]$ is the distortional elastic energy of that state (as in
\eqref{eqn:Fe}), and $C[\nhat]$ is the capacitance of the cell in that
state (given by \eqref{eqn:Cn} for our model problem).  The above
expressions apply more generally than just to our model problem; they
are valid, for example, even if the fields in the cell depend on more
than one space variable (though in that case, one would not have an
explicit formula for $C[\nhat]$, in general).

Stable equilibrium states of the system correspond to minimizers of
the total potential energy $\calG$ with respect to $(\nhat,\DU)$ or
$(\nhat,Q)$, depending on the choice of state variables.  The
capacitance is positive; so
\begin{equation*}
  \min \calG ~ \Rightarrow ~ \DU = V , ~ Q = C V ,
\end{equation*}
and the following minimization problem for $\nhat$ results:
\begin{equation*}
  \min_{\nhat} \Bigl\{ \Fe[\nhat] - \frac12 C[\nhat] V^2 \Bigr\} .
\end{equation*}
Using \eqref{eqn:QDUvsDE} and the equilibrium condition $Q=C[\nhat]V$,
the objective functional above can be written
\begin{equation*}
  \Fe[\nhat] - \frac12 C[\nhat] V^2 =
  \Fe[\nhat] - \frac12 \int_\Omega ( \bmD \cdot \bmE ) \, \dV .
\end{equation*}
Here the terms $\bmD$ and $\bmE$ in the last integral correspond to
the solution of the electrostatics problem in $\Omega$ with the
dielectric tensor associated with the current director field,
${\mathlarger\epstensor}={\mathlarger\epstensor}(\nhat)$, and with a
potential difference of $V$ across the cell.  The minimization with
respect to $\nhat$ is subject to the pointwise constraint $|\nhat|=1$
and appropriate boundary conditions.  The functional
$\Fe - \frac12 \int_\Omega ( \bmD \cdot \bmE ) \, \dV$ is the correct
form of the free energy for equilibrium states of $\nhat$ in an
electric field, as in \eqref{eqn:FnU} (with the identification
$\bmD=\epstensor(\nhat)\bmE$, $\bmE=-\nabla U$).

All of these expressions can be put in explicit forms for our model
problem, for which it is convenient to use the formulation in terms of
$Q$.  The state variables are $\nhat=\nhat(z,t)$ (the director field
in the cell) and $\sigma=\sigma(t)$ (the charge density on the upper
electrode).  Scaling our energies and capacitance by the
cross-sectional area of the cell,
\begin{equation*}
  \tG := \calG / A , \quad \tFe := \Fe / A , \quad \tC := C / A ,
\end{equation*}
and using $Q=\sigma A$, we obtain
\begin{align*}
  \tG[\nhat,\sigma]
  &= \tFe[\nhat] + \frac12 \tC[\nhat]^{-1} \sigma^2 - \sigma V \\
  &= \frac12 \int_0^d \! \bigl( \Knxnz \bigr) \, \dz +
    \frac12 \frac{\,\,\sigma^2}{\,\eps_0} \!
    \int_0^d \! \frac{\dz}{\enxnz} - \sigma V .
\end{align*}
Here we have used \eqref{eqn:Fe}, \eqref{eqn:We}, and \eqref{eqn:Cn}
to provide the expressions for $\Fe[\nhat]$ and $C[\nhat]$.  We shall
work with this mostly in terms of the tilt-angle representation
$\nhat=\cos\theta\,\ex+\sin\theta\,\ez$:
\begin{equation}\label{eqn:calGtilde}
  \tG[\theta,\sigma] =
  \frac12 \int_0^d \bigl(\KCCSS\bigr) \thz^2 \, \dz +
  \frac12 \frac{\,\,\sigma^2}{\,\eps_0} \!
  \int_0^d \! \frac{\dz}{\eCCSS} - \sigma V .
\end{equation}
The associated coupled equilibrium equations from $\delta_\theta\tG=0$
and $\partial\tG/\partial\sigma=0$ are
\begin{gather*}
  \frac{\td}{\dz} \bigl[ \bigl( \KCCSS \bigr) \thz \bigr] =
  \sin\theta \cos\theta \biggl[ (K_3-K_1) \thz^2 -
  \epsa \frac{\,\,\sigma^2}{\,\eps_0} \frac1{(\eCCSS)^2} \biggr] , \\
  \frac{\,\sigma}{\,\eps_0} \int_0^d \! \frac{\dz}{\eCCSS} = V .
\end{gather*}
The latter equation is equivalent to $Q=CV$, which is the correct
equilibrium condition for the charge.  Substituting the equilibrium
value for $\sigma$ above into \eqref{eqn:calGtilde} and into the
equilibrium ODE for $\theta$ above correctly gives the reduced free
energy \eqref{eqn:tFr} and equilibrium equation
\eqref{eqn:thetaODEnonlocal}, in agreement with
\cite[(3.221), (3.226)]{stewart:04}.

\subsection{Dissipation and dynamics}

We wish to use a dissipation principle to obtain a dynamical system
involving the coupled state variables $\nhat$ and $\sigma$.  The
simplest expression for dissipation associated with the dynamics of
the director field is usually given in terms of the single rotational
viscosity parameter $\gamma_1$ via the Rayleigh function
\begin{equation*}
  \frac12 \gamma_1 \! \int_\Omega \,
  \Bigl| \frac{\partial\nhat}{\partial t} \Bigr|^2 \dV =
  \frac12 \gamma_1 A \! \int_0^d
  \Bigl( \frac{\partial\theta}{\partial t} \Bigr)^{\!2} \dz =: \Rtheta .
\end{equation*}
Per unit area, then, we have
\begin{equation*}
  \tRtheta := \frac1{A} \Rtheta = \frac12 \gamma_1 \! \int_0^d
  \Bigl( \frac{\partial\theta}{\partial t} \Bigr)^{\!2} \dz .
\end{equation*}
We have already seen in \eqref{eqn:calRcircuit} that the Rayleigh
dissipation function for the electric circuit is
\begin{equation*}
  \frac12 R \Bigl( \frac{\dQ}{\dt} \Bigr)^{\!2} =
  \frac12 R A^2 \Bigl( \frac{\td\sigma}{\dt} \Bigr)^{\!2} =: \Rsigma ,
\end{equation*}
which leads to
\begin{equation*}
  \tRsigma := \frac1{A} \Rsigma =
  \frac12 R A \Bigl( \frac{\td\sigma}{\dt} \Bigr)^{\!2} .
\end{equation*}
Combining these gives
\begin{equation}\label{eqn:calRtilde}
  \tR := \tRtheta + \tRsigma = \frac12 \gamma_1 \! \int_0^d
  \Bigl( \frac{\partial\theta}{\partial t} \Bigr)^{\!2} \dz +
  \frac12 R A \Bigl( \frac{\td\sigma}{\dt} \Bigr)^{\!2} ,
\end{equation}
which is the form we shall adopt for our combined dissipation
function.

The total potential energy \eqref{eqn:calGtilde} and Rayleigh function
\eqref{eqn:calRtilde} produce the correct circuit dynamics:
\begin{equation*}
  \frac{\partial\tG}{\partial\sigma} +
  \frac{\partial\tR}{\partial\dot{\sigma}} = 0 ~ \Rightarrow ~
  \frac{\,\sigma}{\,\eps_0} \int_0^d \! \frac{\dz}{\eCCSS} -
  V + R A \frac{\td\sigma}{\dt} = 0 ~ \Rightarrow ~
  R \frac{\dQ}{\dt} + \frac1{C} Q = V ,
\end{equation*}
using \eqref{eqn:Cn} and $Q = \sigma A$---this agrees with
\eqref{eqn:QODEIVP}.  The dynamics for the director angle results from
the analogous expression for $\theta$:
\begin{equation*}
  \delta_\theta\tG + \delta_{\dot{\theta}}\tR = 0 ~ \Rightarrow ~
  \frac{\partial W}{\partial\theta} - \frac{\partial}{\partial z}
  \Bigl( \frac{\partial W}{\partial\thz} \Bigr) +
  \gamma_1 \frac{\partial\theta}{\partial t} = 0 ,
\end{equation*}
with $W$ the free-energy density that results from combining the
integrated terms in \eqref{eqn:calGtilde}.  In expanded form, the PDE
for $\theta$ is given by
\begin{subequations}\label{eqn:coupled-dynamical-system}
\begin{multline}\label{eqn:dthetadt}
  \gamma_1 \frac{\partial\theta}{\partial t} =
  \frac{\partial}{\partial z}
  \Bigl[ \bigl( \KCCSS \bigr) \frac{\partial\theta}{\partial z} \Bigr] \\
  {} - \sin\theta \cos\theta \biggl[ (K_3-K_1)
  \Bigl( \frac{\partial\theta}{\partial z} \Bigr)^{\!2} -
  \epsa \frac{\,\,\sigma^2}{\,\eps_0} \frac1{(\eCCSS)^2} \biggr] .
\end{multline}
Our dynamical system, then, consists of the PDE \eqref{eqn:dthetadt}
above for $\theta$, supplemented by auxiliary conditions
\begin{equation}
  \theta(0,t) = \theta(d,t) = 0 , \quad \theta(z,0) = \theta_0(z) ,
\end{equation}
and the ODE IVP for $\sigma$
\begin{equation}\label{eqn:dsigmadt}
  R A \frac{\td\sigma}{\dt} +
  \frac{\,\sigma}{\,\eps_0} \int_0^d \! \frac{\dz}{\eCCSS} = V , \quad
  \sigma(0) = \sigma_0 ,
\end{equation}
\end{subequations}
with the initial states $\theta_0$ and $\sigma_0$ prescribed.
The variational approach used here to obtain the director dynamics
equation \eqref{eqn:dthetadt} is similar to that given in
\cite[\S4.3]{stewart:04}---compare \eqref{eqn:dthetadt} with
\cite[(4.152)]{stewart:04}.  See also \cite[\S5.9]{stewart:04},
where the same ideas are used to model the dynamics of certain \Freed\
transitions.

Some of the terms above can be related to more familiar expressions.
The term involving $\sigma^2$ in \eqref{eqn:dthetadt} corresponds to
the ``dielectric torque'' $\bmD\times\bmE$, the couple per unit volume
exerted by the electric field on the director field.  In our system,
as in \eqref{eqn:DepsE} and \eqref{eqn:EUzDzsigma}, we have
\begin{equation*}
  \bmE = E \ez , \quad
  \bmD = \epstensor(\nhat) \bmE =
  \eps_0 E \bigl[ \epsa n_x n_z \ex + \bigl( \enxnz \bigr) \ez \bigr] .
\end{equation*}
In terms of $\theta$, then,
\begin{equation*}
  \bmD \times\bmE = - \eps_0 \epsa E^2 \sin\theta \cos\theta \, \ey .
\end{equation*}
The connection between $E^2$ and $\sigma^2$ follows from
\begin{equation*}
  \sigma = - D_z ~ \Rightarrow ~
  \sigma^2 = \eps_0^2 E^2 ( \eCCSS )^2 ,
\end{equation*}
which gives
\begin{equation*}
  \bmD \times \bmE = - \epsa \frac{\,\,\sigma^2}{\,\eps_0}
  \frac{\sin\theta\cos\theta}{(\eCCSS)^2} \ey ,
\end{equation*}
as in \eqref{eqn:dthetadt}.  Also, using \eqref{eqn:Cn} we have
\begin{equation*}
  C[\theta]^{-1} = \frac1{\eps_0 A} \int_0^d \! \frac{\dz}{\eCCSS} ,
\end{equation*}
and \eqref{eqn:dsigmadt} can be written
\begin{equation*}
  R A \frac{\td\sigma}{\dt} + C[\theta]^{-1} \! A \sigma = V
  ~~ \text{or} ~~
  R \frac{\td Q}{\dt} + C[\theta]^{-1} Q = V ,
\end{equation*}
as in \eqref{eqn:QODEIVP}.

As we have already noted (and as we shall see in
\S\ref{sec:TimeScales}), the time scale for \eqref{eqn:dsigmadt} is
often faster than that for \eqref{eqn:dthetadt}.  If one were to
choose to model the charge distribution as adjusting instantaneously
to changes in the director field, then one would have
$\td\sigma/\dt=0$ in \eqref{eqn:dsigmadt}, and that equation would
collapse to the equilibrium condition for $\sigma$ in the previous
section.  Substituting the equilibrium value for $\sigma$ into
\eqref{eqn:dthetadt} would give the director-dynamics equation for the
reduced free energy $\tFr$ in \eqref{eqn:tFr} (with $\DU=V$), that is,
the dynamical version of \eqref{eqn:thetaODEnonlocal} (again with
$\DU=V$).

\subsection{Time-varying $\bmD$ versus $\bmE$}

With an electric field that varies in space and in time
($\bmE = \bmE(\bmx,t)$), the linear relationship between $\bmD$ and
$\bmE$ is in general nonlocal in space and in time; however, for
macroscopic models of pure dielectrics, nonlocality in space can
generally be ignored \cite[\S I.4]{jackson:75}, \cite[\S77, \S78,
\S103]{landau:lifshitz:pitaevskii:93}.  Nonlocality in time can be an
issue in some circumstances, such as fast-switching experiments with
electric fields from electric pulses of large voltage and short
duration \cite{gu:yin:shiyanovskii:lavrentovich:07}.  If the
dielectric properties of the medium remain constant, spatial
dispersion is ignored, and the electric field is time harmonic, then
the induced polarization will be time harmonic with the same frequency
and hence so will $\bmD$ (via $\bmD=\eps_0\bmE+\bmP$):
\begin{equation*}
  \bmE(\bmx,t) = e^{-i\omega_0t} \bmE_0(\bmx) ~~ \Rightarrow ~~
  \bmD(\bmx,t) = e^{-i\omega_0t} \epstensor(\bmx,\omega_0) \bmE_0(\bmx) .
\end{equation*}
See \cite[\S I.4]{jackson:75},
\cite[\S77]{landau:lifshitz:pitaevskii:93}.  This includes the special
case of a static electric field ($\omega_0=0$):
$\bmD(\bmx)={\mathlarger\epstensor}(\bmx)\bmE(\bmx)$.

The relationship between $\bmD$ and $\bmE$ in the time-harmonic case
is \emph{frequency dependent}, however
(${\mathlarger\epstensor}={\mathlarger\epstensor}(\bmx,\omega)$), and
so for a general time-varying electric field (again in a medium of
constant dielectric properties and no spatial dispersion), one has
\begin{equation}\label{eqn:DEnonlocal}
  \bmD(\bmx,t) = \eps_0 \bmE(\bmx,t) + \eps_0 \! \int_{-\infty}^t \!
  \alphatensor(\bmx,t-t') \bmE(\bmx,t') \, \td t' ,
\end{equation}
as written in \cite{gu:yin:shiyanovskii:lavrentovich:07}.  The tensor
field ${\mathlarger\alphatensor}$ comes from the inverse Fourier
transform of the dispersion relation, that is,
\begin{equation*}
  \epstensor(\bmx,\omega) = \eps_0 \Bigl[ \bfI + \! \int_0^{\infty} \!
  e^{i\omega t} \alphatensor(\bmx,t) \, \dt \Bigr] .
\end{equation*}
An illustrative example (for a homogeneous, isotropic medium) is given
in \cite[\S7.10]{jackson:75}, where a ``one-resonance dispersion
model'' is employed (a complex rational function of $\omega$).  The
poles of the dispersion model are in the lower half of the complex
$\omega$ plane, which gives rise to a one-sided inverse Fourier
transform and an appropriately causal relationship of the form
\eqref{eqn:DEnonlocal}.  From a physical point of view, at high
frequencies, changes in the induced polarization can lag changes in
the electric field.  An electric pulse resolves into Fourier modes of
arbitrarily high frequency, and because of this, these
nonlocal-in-time effects (``dielectric relaxation'') can become
important.  In \cite{gu:yin:shiyanovskii:lavrentovich:07} and
\cite{shiyanovskii:lavrentovich:10}, the term ``Dielectric Memory
Effects'' is used to describe them.

The dynamics of liquid-crystal systems induced by time-varying
electric fields is even more complicated, since in that case, the
dielectric properties of the medium are changing in time as well, and
the convolution representation \eqref{eqn:DEnonlocal} is no longer
valid.  A research program described in the review article
\cite{shiyanovskii:lavrentovich:10} (which contains references to
earlier works) addressed this issue and developed a generalization of
\eqref{eqn:DEnonlocal} to the case of a time-varying director field
$\nhat=\nhat(\bmx,t)$.  The theory was used to model successfully the
experiments with fast-switching (pulse-driven) dynamics of
\cite{takanashi:maclennan:clark:98} and others.

These issues are beyond the scope of our investigation here.  Our
assessment is that the local relationship
$\bmD={\mathlarger\epstensor}\bmE$ is valid in static equilibrium and
more generally if $\bmE$ is time harmonic and the dielectric
properties of the medium are constant.  The relationship
$\bmD={\mathlarger\epstensor}\bmE$ is a good approximation if $\bmE$
contains only low-frequency content and the director dynamics are
relatively slow.  For our purposes here, we accept and adopt the local
relationship $\bmD={\mathlarger\epstensor}\bmE$ and acknowledge the
limitations and shortcomings of that in some settings.

\subsection{Time scales}\label{sec:TimeScales}

The dynamic processes associated with a system such as the one under
consideration here exhibit several different time scales, including
those for director dynamics, circuit dynamics, and the evolution of
the electric field in the cell.  The time scales for the changes in
the circuit and for the dynamics of the director field can be gleaned
from rescalings of \eqref{eqn:dsigmadt} and \eqref{eqn:dthetadt}.
First, it is convenient to relate the integral expression in
\eqref{eqn:dsigmadt} to the capacitance \eqref{eqn:Cn} as follows:
\begin{equation*}
  C[\theta] = \eps_0 A \biggl[
  \int_0^d \!\! \frac{\dz}{\eCCSS} \biggr]^{-1} =
  \frac{\eps_0 A}{d} \epsr[\theta] ,
\end{equation*}
where $\epsr[\theta]$ is the \emph{effective relative dielectric
  constant} of the cell, given by
\begin{equation*}
  \epsr[\theta]^{-1} = \frac1d \int_0^d \!\! \frac{\dz}{\eCCSS} .
\end{equation*}
This quantity is dimensionless, $O(1)$, and satisfies
\begin{equation*}
  \eperp \le \epsr[\theta] \le \epara ,
\end{equation*}
since $\eperp < \epara$ for our system.  In terms of it,
\eqref{eqn:dsigmadt} can be written
\begin{subequations}\label{eqn:partially_scaled}
\begin{equation}
  \tausig \frac{\td\sigbar}{\dt} + \epsr[\theta]^{-1} \sigbar = 1 , \quad
  \tausig := R \frac{\eps_0 A}{d} , \quad
  \sigbar := \frac{\sigma}{\eps_0 V / d} .
\end{equation}
Note that $\eps_0 A / d$ would be the capacitance of the cell and
$\eps_0 V/d$ would be the surface charge density on the upper
electrode if there were a vacuum between the electrodes.  Note also
that in steady state, we have $\td\sigbar / \dt = 0$ and
$\sigbar = \epsr[\theta]$.

For a fairly typical experimental setup, the ``RC load'' $\tausig$ is
generally in the sub-microsecond range.  For example, with
$R=100\,\Omega$, $\eps_0=8.854\times10^{-12}\,\text{F/m}$,
$A=10\,\text{cm}^2$, and $d=5\,\mu\text{m}$, we obtain
$\tausig \doteq 1.77\times10^{-7}\,\text{s}$.  In
\cite{jang:clark:01}, $\tausig$ is estimated to be $2\,\mu\text{s}$
and is a limiting factor in the experiment presented there.  In the
fast-switching experiments discussed in
\cite{gu:yin:shiyanovskii:lavrentovich:07} and
\cite{takanashi:maclennan:clark:98}, measures are taken to minimize
$\tausig$ (including the use of gold connectors to minimize resistance
and cells of small area to minimize capacitance), and values for
$\tausig$ in the nanosecond range are reported.

If one scales $z$ by the cell gap ($\zbar=z/d$), then
\eqref{eqn:dthetadt} can be put in the form
\begin{multline}
  \frac{\partial\theta}{\partial t} = \frac1{\,\tauK} \biggl\{
  \frac{\partial}{\partial\zbar} \Bigl[
  (\Kbar_1\CC+\Kbar_3\SS) \frac{\partial\theta}{\partial\zbar} \Bigr] -
  \sin\theta \cos\theta (\Kbar_3-\Kbar_1)
  \Bigl( \frac{\partial\theta}{\partial\zbar} \Bigr)^{\!2} \biggr\} \\
  {} + \frac1{\,\tauV} \sigbar^2
  \frac{\sin\theta \cos\theta}{(\eCCSS)^2} ,
\end{multline}
where
\begin{equation}
  \tauK := \frac{\gamma_1d^2}{K} , \quad
  \tauV := \frac{\gamma_1d^2}{\eps_0\epsa V^2} , \quad
  \Kbar_1 := \frac{\,K_1}{K} , \quad \Kbar_3 := \frac{\,K_3}{K} ,
\end{equation}
\end{subequations}
with $K$ a representative value for $K_1$ and $K_3$.  This exposes two
times scales associated with director reorientation: $\tauK$ and
$\tauV$.  These correspond to ``switch off'' and ``switch on'' times,
which are usually written
\begin{equation*}
  \tauoff = \frac{\gamma_1d^2}{K\pi^2} , \quad
  \tauon = \frac{\gamma_1}{\eps_0\epsa E^2} .
\end{equation*}
See \cite[\S5.9]{stewart:04}.

The switch-off time $\tauoff$ is the time scale for the slowest
decaying mode of $\gamma_1 \theta_t = K \theta_{zz}$, $\theta(0) =
\theta(d) = 0$.  It gives the time it takes for a distorted director
field to relax back to its ground state under the influence of only
distortional elastic forces.  This is a relatively slow process,
usually estimated to be in the range of 10s of milliseconds for
typical cells and materials (e.g., with $\gamma_1=0.1\,\text{Pa\,s}$,
$d=5\,\mu\text{m}$, and $K=10\,\text{pN}$, we obtain $\tauoff \doteq
2.53\times10^{-2}\,\text{s}$).

The switch-on time $\tauon=\tauV$ (with $E=V/d$) corresponds to the
time it takes to align the director field with an applied electric
field.  It can be made quite small by using large voltages.  For
example, with $\gamma_1=0.1\,\text{Pa\,s}$, $d=5\,\mu\text{m}$,
$\eps_0=8.854\times10^{-12}\,\text{F/m}$, $\epsa=10$, and
$V=100\,\text{volts}$, we have
$\tauV \doteq 2.82 \times 10^{-6}\,\text{s}$.  For small voltages near
the \Freed\ threshold, on the other hand (e.g., with
$V=1\,\text{volt}$), we have
$\tauV \doteq 2.82 \times 10^{-2}\,\text{s}$.  In the fast-switching
experiments in \cite{geis:lyszczarz:osgood:kimball:10,
  gu:yin:shiyanovskii:lavrentovich:07,takanashi:maclennan:clark:98},
using cells of narrow gaps and electrical pulses of 100s of volts,
$\tauV$\!'s of the order of 10s of nanoseconds are reported.

By comparison, the evolution of the electric field in the cell is
governed by the time-dependent Maxwell equations, and the time scale
associated with this type of wave equation (which we denote $\tauE$)
corresponds to the width of the cell gap divided by the speed of light
in the medium.  For a cell gap $d=10\,\mu\text{m}$ with relative
dielectric permittivity $\epsr=10$ and relative magnetic permeability
$\mur=1$, we obtain $\tauE \doteq 1.05 \times 10^{-13}\,\text{s}$,
which is four orders of magnitude faster than the smallest
$\tausig$\!'s and $\tauV$\!'s we have found in published results on
experiments on fast switching of liquid crystal cells.  This justifies
treating the electric field as adjusting instantaneously to any
changes in the circuit and cell.

Thus, for the kinds of experimental systems that we have in mind, we
consider time scales in the following ranges:
\begin{equation*}
  10^{-9}\,\text{s} \lesssim \tausig \lesssim 10^{-6}\,\text{s} , \quad
  \tauK \approx 10^{-2}\,\text{s} , \quad
  10^{-8}\,\text{s} \lesssim \tauV \lesssim 10^{-2}\,\text{s} , \quad
  \tauE \approx 10^{-13}\,\text{s} .
\end{equation*}
In the numerical examples discussed in the next section, we have taken
$\tausig=10^{-6}\,\text{s}$, $\tauK=10^{-2}\,\text{s}$, and
$\tauV=10^{-7}\,\text{s}$, $10^{-6}\,\text{s}$, and
$10^{-5}\,\text{s}$, for purposes of illustration, with the electric
field taken to adjust instantaneously ($\tauE=0$, in essence, as we
have assumed throughout).

\section{Numerical illustration}\label{sec:numerics}

We illustrate the coupled dynamics of
\eqref{eqn:coupled-dynamical-system} by the numerical modeling of a
simple ``switch on'' experiment.  For simplicity, we assume equal
elastic constants: $K_1=K_3=K$.  Starting from the partially scaled
system \eqref{eqn:partially_scaled}, we express times in units of
$\tausig$ to obtain
\begin{subequations}\label{eqn:fully_scaled}
\begin{gather}
  \label{eqn:fully_scaled_a}
  \frac{\td\sigbar}{\dtbar} + \epsr[\theta]^{-1} \sigbar = 1 , \quad
  \epsr[\theta]^{-1} = \int_0^1 \!\! \frac{\dzbar}{\eCCSS} , \\
  \label{eqn:fully_scaled_b}
  \frac{\partial\theta}{\partial\tbar} =
  \frac1{\,\tauKbar} \frac{\partial^2\theta}{\partial\zbar^2} +
  \frac1{\,\tauVbar} \sigbar^2
  \frac{\sin\theta \cos\theta}{(\eCCSS)^2} , \quad 0 < \zbar < 1 ,
\end{gather}
where
\begin{equation*}
  \tbar := \frac{t}{\,\tausig} , \quad
  \tauKbar := \frac{\tauK}{\,\tausig} , \quad
  \tauVbar := \frac{\tauV}{\,\tausig} .
\end{equation*}
We add a small pretilt to the boundary conditions on $\theta$, in
order to bias the director to rotate counter clockwise when the
electric field is switched on, and we take the initial state of the
director field to be parallel to this.  There is assumed to be no
excess charge on the electrodes when the circuit is closed at time
$\tbar=0$.  The boundary and initial conditions are thus given by
\begin{equation}
  \theta(0,\tbar) = \theta(1,\tbar) = 0.1 , \quad
  \theta(\zbar,0) = 0.1 , \quad
  \sigbar(0) = 0 .
\end{equation}
\end{subequations}

For the relative dielectric permittivities, we use $\eperp=5$ and
$\epara=15$, which are comparable to the values for the typical liquid
crystal 5CB---see \cite[Appendix D, Table\,D.3]{stewart:04}.  Based
upon the discussion of time scales in the previous section, we choose
the following values for our numerical experiment:
\begin{equation*}
  \tauK = 10^{-2}\,\text{s} , \quad
  \tausig = 10^{-6}\,\text{s} , \quad
  \tauV = 10^{-7}\,\text{s} , ~ 10^{-6}\,\text{s} , ~ 10^{-5}\,\text{s} , 
\end{equation*}
giving
\begin{equation*}
  \tauKbar = 10^4 , \quad
  \tauVbar = 10^{-1} , ~ 10^0 , ~ 10^1 .
\end{equation*}
The three values of $\tauVbar$ cover the situations when the director
switching dynamics are faster than, equal to, and slower than the
circuit dynamics.

\begin{figure}
  \centering
  \subfloat[]{\label{fig:tauV_pointone_a}
    \includegraphics[width=.49\linewidth]{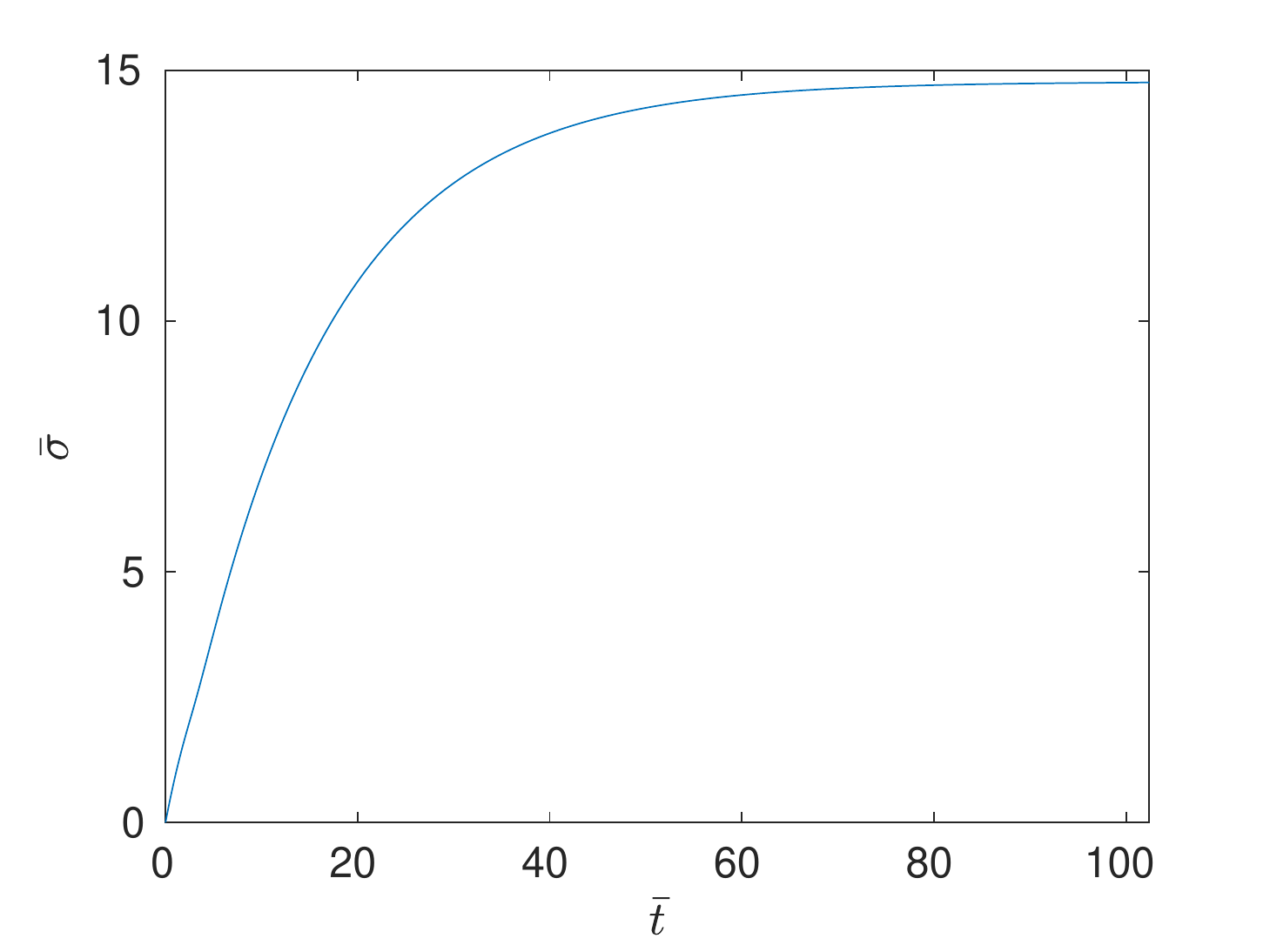}}
  \subfloat[]{\label{fig:tauV_pointone_b}
    \includegraphics[width=.49\linewidth]{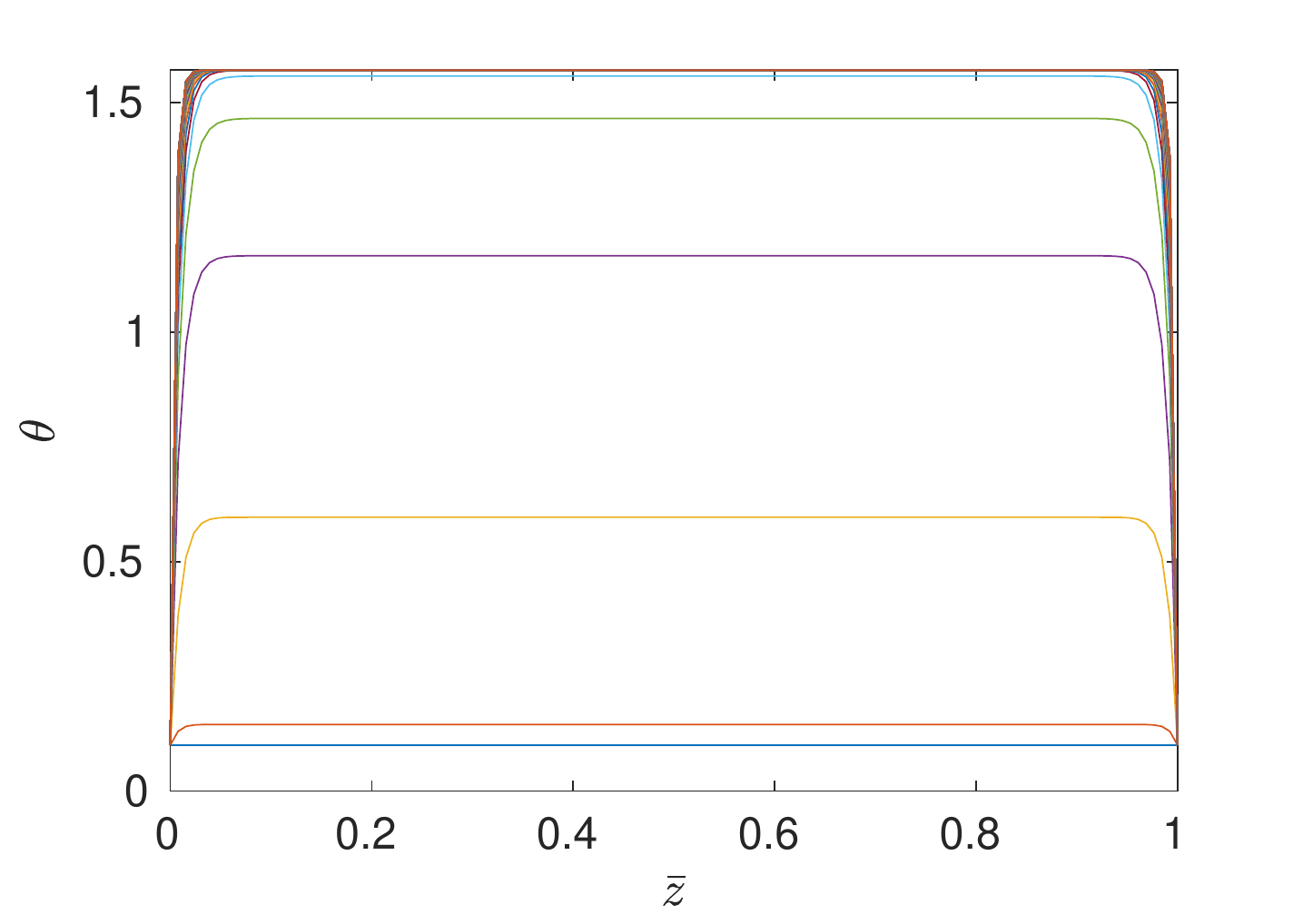}} \\
  \subfloat[]{\label{fig:tauV_one_a}
    \includegraphics[width=.49\linewidth]{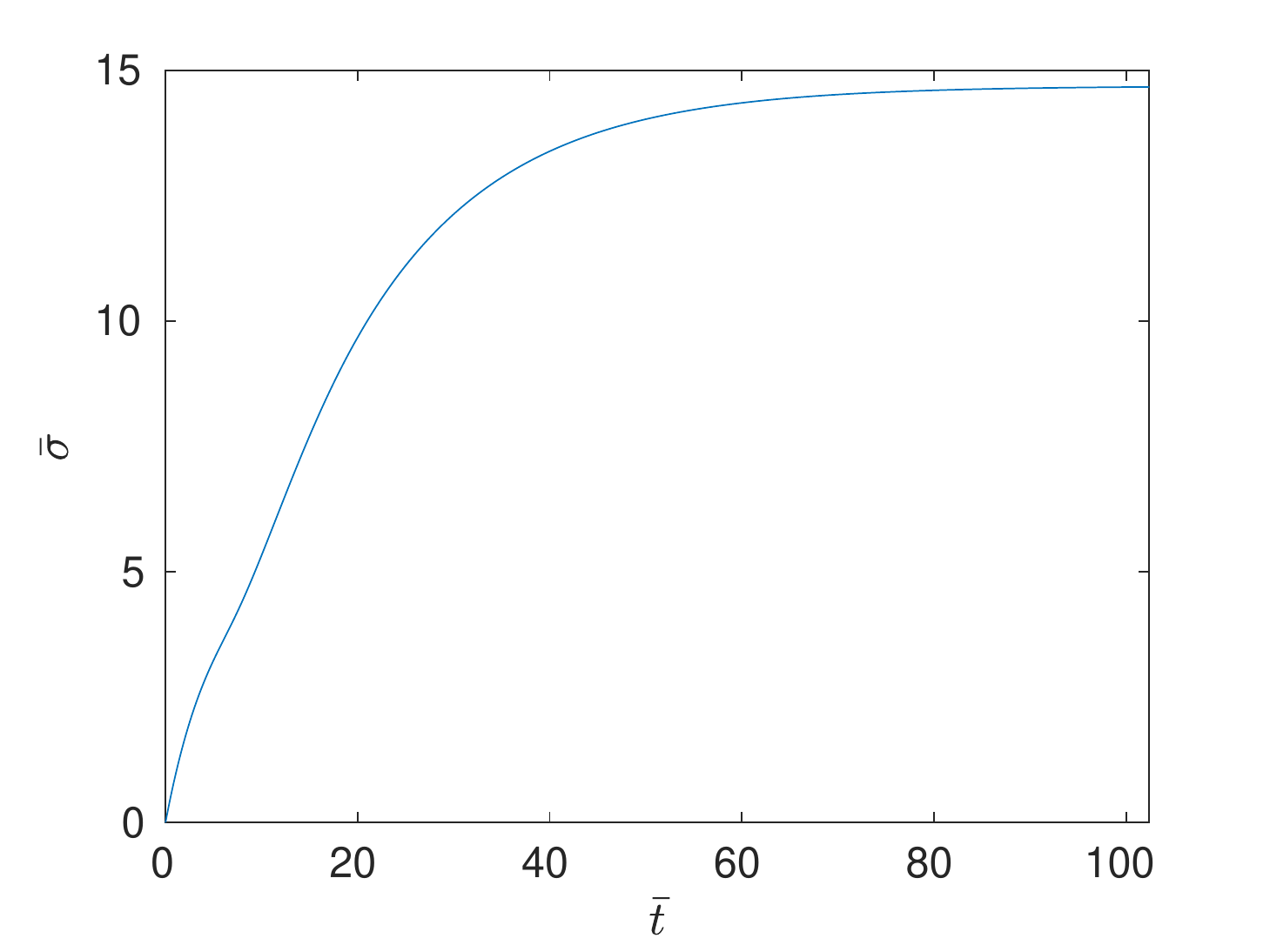}}
  \subfloat[]{\label{fig:tauV_one_b}
    \includegraphics[width=.49\linewidth]{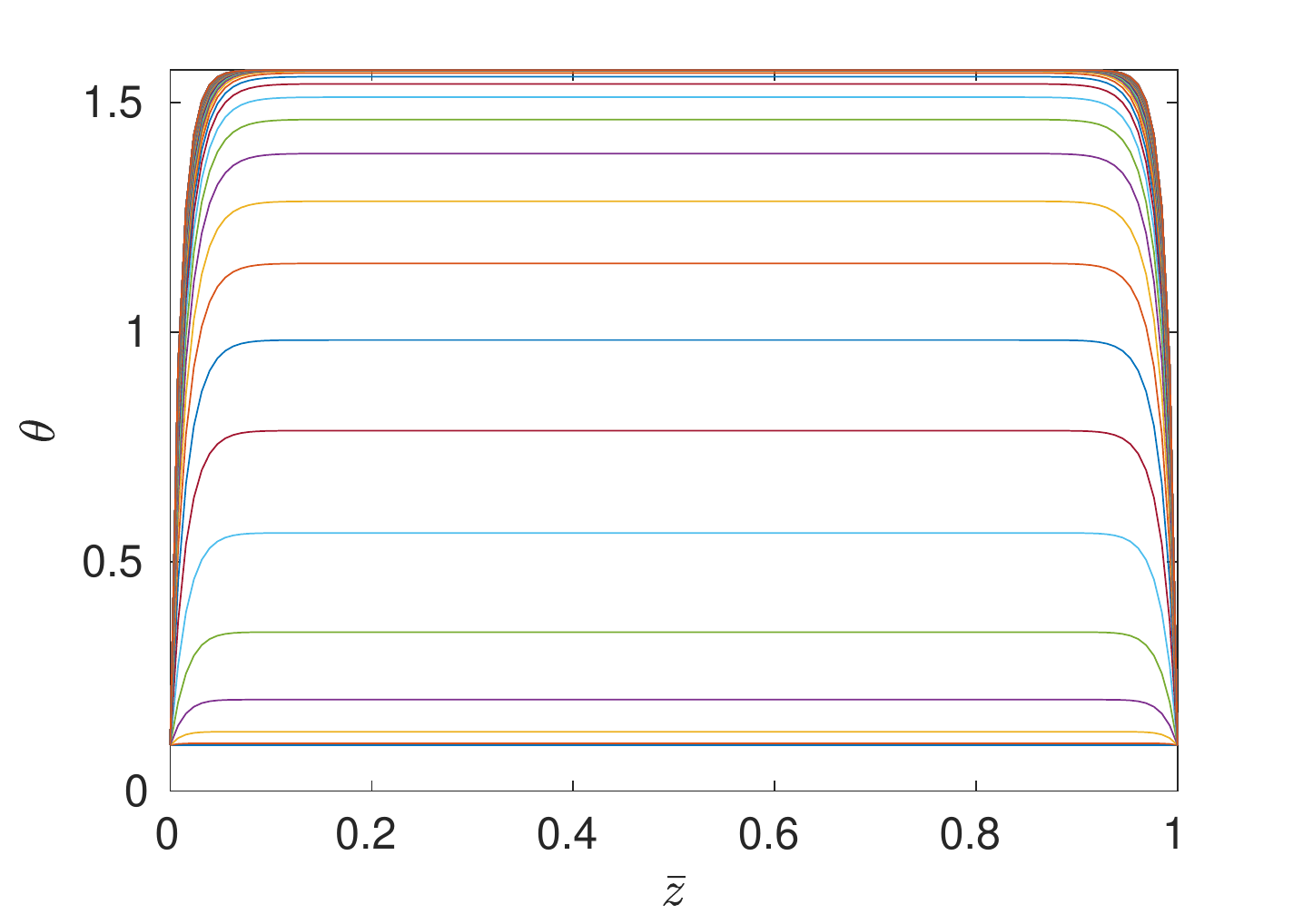}} \\
  \subfloat[]{\label{fig:tauV_ten_a}
    \includegraphics[width=.49\linewidth]{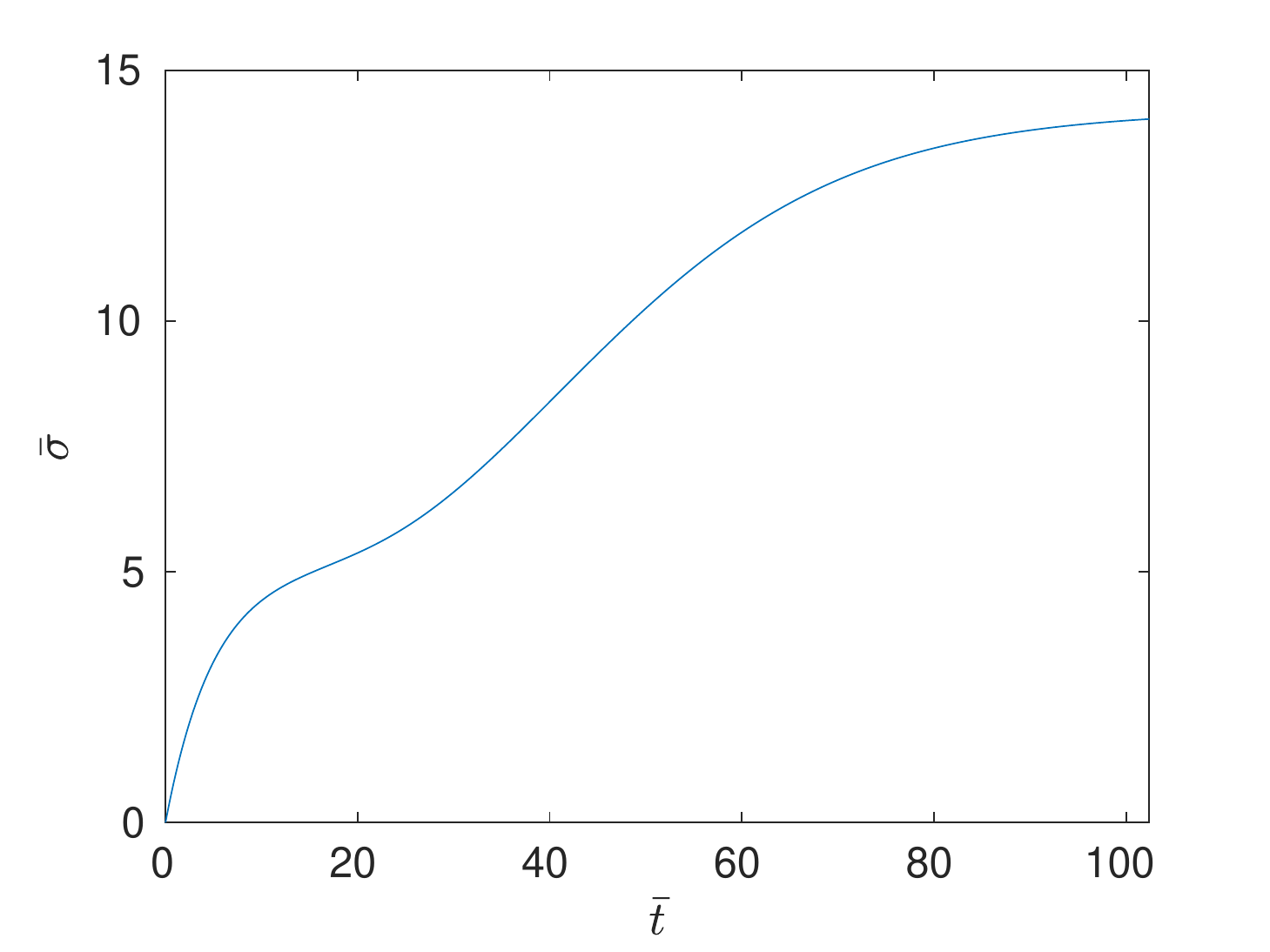}}
  \subfloat[]{\label{fig:tauV_ten_b}
    \includegraphics[width=.49\linewidth]{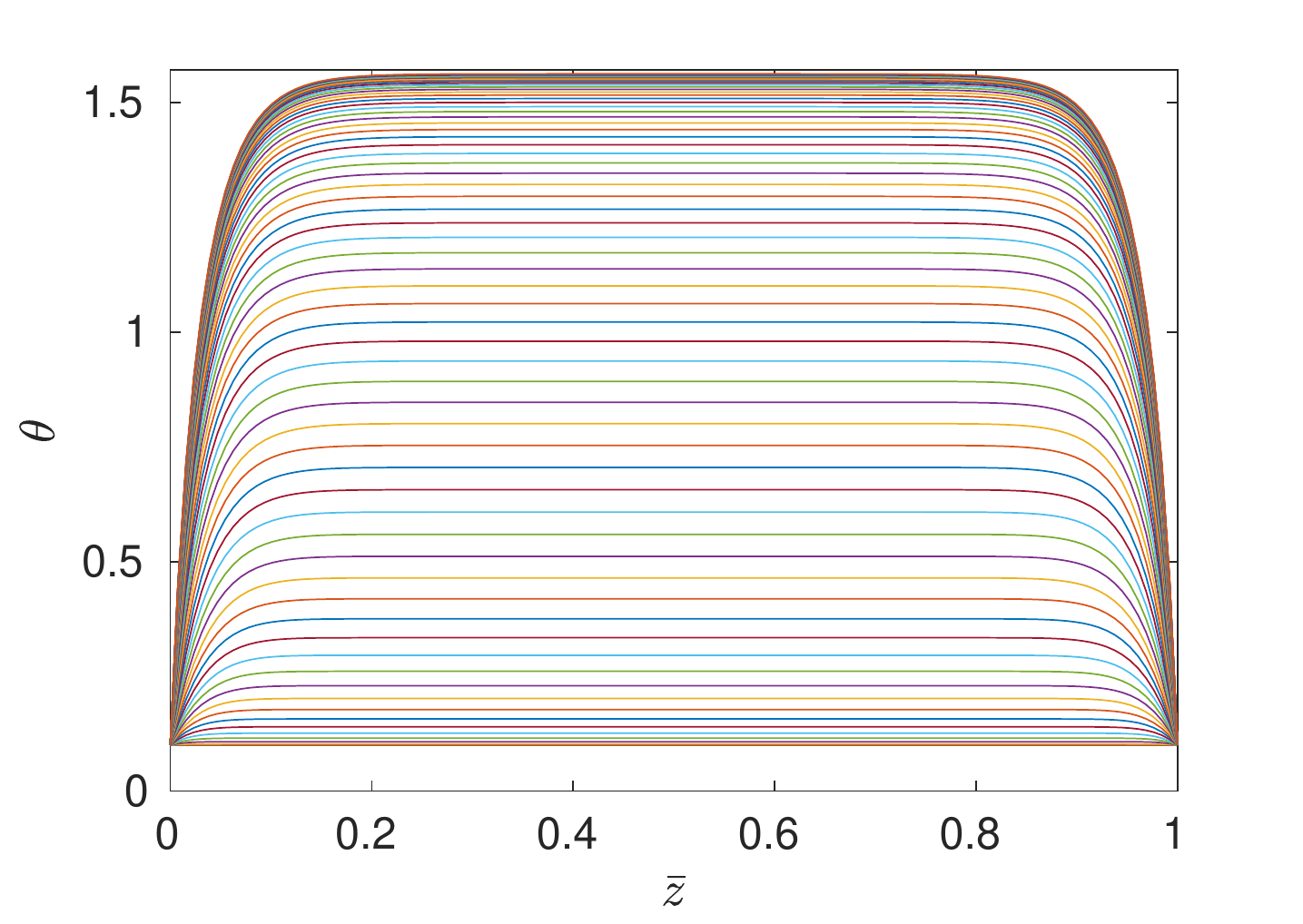}}
  \caption{Coupled dynamics \eqref{eqn:fully_scaled}.
    Figures\,\ref{fig:tauV_pointone_a}, \ref{fig:tauV_one_a},
    \ref{fig:tauV_ten_a}: electrode charge density versus
    time ($\sigbar=\sigma/(\eps_0 V/d)$, $\tbar=t/\tausig$).
    Figures\,\ref{fig:tauV_pointone_b}, \ref{fig:tauV_one_b},
    \ref{fig:tauV_ten_b}: time snapshots of director tilt
    angle $\theta$ versus position ($\zbar=z/d$, 16 time steps per
    snapshot, time step $\Delta\tbar=0.1$).  Parameters: $\eperp=5$,
    $\epara=15$, $\tauKbar=\tauK/\tausig=10^4$,
    $\tauVbar=\tauV/\tausig=10^{-1}$
    (figures\,\ref{fig:tauV_pointone_a}, \ref{fig:tauV_pointone_b}),
    $10^0$ (figures\,\ref{fig:tauV_one_a}, \ref{fig:tauV_one_b}),
    $10^1$ (figures\,\ref{fig:tauV_ten_a}, \ref{fig:tauV_ten_b}).}
  \label{fig:sigma_t_theta_z}
\end{figure}

Standard finite differences were employed to discretize our model
\eqref{eqn:fully_scaled}: explicit Euler for
\eqref{eqn:fully_scaled_a}, Forward Time Centered Space (FTCS) for
\eqref{eqn:fully_scaled_b}.  The same time step was used for both
equations ($\Delta\tbar=0.1$, 10 times steps per unit $\tausig$), with
a spatial grid of 128 uniform cells in $\zbar$.  A Composite Trapezoid
Rule (on the same spatial grid) was used to approximate the integral
in the functional $\epsr[\theta]$ in \eqref{eqn:fully_scaled_a}.  For
each of the values $\tauVbar=10^{-1}$, $10^0$, and $10^1$, $\sigbar$
is plotted against $\tbar$, along with ``time snapshots'' of $\theta$
versus $\zbar$ for every 16th time step.  In all three cases, the same
number of time steps per snapshot (16) and the same number of total
time steps (1024) were used, in order to facilitate comparison.  The
results are presented in figure\,\ref{fig:sigma_t_theta_z}.

The results are as one would anticipate.  The case $\tauVbar=10^{-1}$
corresponds to a high voltage and fast switching: the director quickly
saturates (after 5--6 snapshots), and the charge density behaves
accordingly ($\sigbar \approx \epara ( 1 - e^{-\tbar} )$).  We note
that in the steady state limit $\tbar \rightarrow \infty$, we have
$\sigbar \rightarrow \epsr[\theta_{\infty}] \approx \epara$, since
$\theta_{\infty} \approx \pi/2$ (except for boundary layers near
$\zbar=0$ and $\zbar=1$).  The case $\tauVbar=10^0$ corresponds to a
moderate voltage and switching ($\tauV=\tausig$): $\theta$ saturates
after 15--16 snapshots, and one starts to see an inflection in
$\sigbar$.  The last case, $\tauVbar=10^1$, is associated with a low
voltage and slow switching: it takes 50--60 snapshots for the director
to align with the electric field, and one sees a pronounced inflection
in $\sigbar$ versus $\tbar$.  This is because the time scale for the
$\sigbar$ dynamics is approximately $R \eps_0 \eperp A / d$ in the
early stages (when $\theta \approx 0$ and
$\epsr[\theta] \approx \eperp$), but the time scale is approximately
$R \eps_0 \epara A / d$ in the later stages (when
$\theta \approx \pi/2$ and $\epsr[\theta] \approx \epara$).  Since
$\eperp=5$ and $\epara=15$, these time scales differ by a factor of
three, and this leads to the change in the rate of approach of
$\sigbar$ to its limiting value.  We emphasize that this is merely an
illustration of the coupled dynamics that emerge from our simple
model; in order to model carefully the dynamics in an actual
fast-switching experiment, for example, one would need to take into
account other influences, such as ``dielectric relaxation''
\cite{gu:yin:shiyanovskii:lavrentovich:07,shiyanovskii:lavrentovich:10}.

\section{Conclusions}\label{sec:conclusions}

We have modeled a nematic-liquid-crystal cell subject to an electric
field created by electrodes held at constant potential as a variable
capacitor in an RC circuit.  The general model couples the state of
the liquid-crystal director field $\nhat$ in the cell with the state
of the electric circuit, characterized in terms of either the total
charge on the upper electrode, $Q$, or the potential difference
between the electrodes, $\DU$.  A dynamical system was derived for an
example in the splay-\Freed\ geometry, subject to several simplifying
assumptions: no fluid flow in the cell, an electric field that adjusts
instantaneously to changes in $\nhat$, fields in the cell that are
functions of one space variable only, and a single rotational
viscosity for energy dissipation associated with
$\partial\nhat/\partial t$.  The dynamical system, given in
\eqref{eqn:coupled-dynamical-system}, involves a PDE for director
dynamics ($\nhat=\nhat(z,t)$) coupled to an ODE for charge dynamics
($\sigma=\sigma(t)$, where $\sigma$ is the density of free charge on
the surface of the upper electrode).  We have produced estimates for
the time scales of the various dynamic processes and have provided
numerical examples illustrating the coupled dynamics for three
different cases relating the time scale for director dynamics to that
of the circuit dynamics.  We have made an effort to show consistency
with established results, where possible.

The original motivation for this exercise was to understand better the
dynamical characterization of local stability of equilibrium states of
such systems, which can be characterized as stationary points of a
free-energy functional of the form \eqref{eqn:FnU}:
\begin{equation*}
  \calF[\nhat,U] = \int_\Omega \Bigl[ \We(\nhat,\nabla\nhat) -
  \frac12 \epstensor(\nhat) \nabla U \cdot \nabla U \Bigr] \, \dV .
\end{equation*}
Here ${\mathlarger\epstensor}$ is the (positive definite) dielectric
tensor and $U$ is the electric potential (related to the electric
field $\bmE$ via $\bmE=-\nabla U$).  The minimax nature of the
critical points of $\calF$ is at odds with the expected picture of an
out-of-equilibrium pair $(\nhat,U)$ relaxing to a locally stable state
by minimizing free energy.  This confusion, as we have seen from our
analysis, stems from the fact that the free-energy functional above
presumes either a static electric field or an electric field that
adjusts instantaneously to changes in $\nhat$.

The more primitive expression for the potential energy of the system
is $\calG$ as in \eqref{eqn:calG}:
\begin{equation*}
  \calG = \Fe + \Ecap + \Ebat = \Fe + \frac12 Q \DU - Q V .
\end{equation*}
Here $\Fe$ is the distortional elasticity, as in \eqref{eqn:Fe} and
\eqref{eqn:We}, and $V$ is the voltage of the battery.  The
combination $\Fe+\Ecap$ gives the potential energy of the cell
(associated with the work done in distorting the director field plus
work done in moving charge on/off the electrodes); while $\Ebat$ is
the potential energy associated with the battery.  The total potential
energy for the system can be expressed in different forms depending
the choice of state variable for the circuit, as in \eqref{eqn:GnDU}
and \eqref{eqn:GnQ}:
\begin{equation*}
  \calG[\nhat,\DU] = \Fe[\nhat] +
  C[\nhat] \Bigl[ \frac12 (\DU)^2 - V \DU \Bigr] , \quad
  \calG[\nhat,Q] = \Fe[\nhat] + \frac12 C[\nhat]^{-1} Q^2 - V Q .
\end{equation*}
In either case, equilibria are locally minimizing with respect to the
pair $(\nhat,\DU)$ or $(\nhat,Q)$.

In order to transform $\calG$ into the form $\calF$ above, one must
assume (1)~either a static electric field or an electric field that
adjusts instantaneously to any changes in $\nhat$ (so that
$\curl\bmE=\bfzero$ and $\bmE=-\nabla U$) and (2)~either equilibrium
conditions in the circuit (no current) or a circuit that adjusts
instantaneously to any changes in the capacitance of the cell (so that
$\DU=V$ and $Q=CV$ at all times).  In
\S\ref{sec:system-potential-energy}, we have described how $\calG$
above can be transformed to $\calF$ if one makes these assumptions
(and also employs the constitutive assumption
$\bmD={\mathlarger\epstensor}(\nhat)\bmE$), the main points being that
$\bmE=-\nabla U$ and $\div\bmD=0$ imply
\begin{equation*}
  Q \DU = \int_\Omega ( \bmD \cdot \bmE ) \, \dV ,
\end{equation*}
$\DU=V$ gives
\begin{equation*}
  \Ecap + \Ebat = \frac12 Q V - Q V = - \frac12 Q V =
  - \frac12 \int_\Omega ( \bmD \cdot \bmE ) \, \dV ,
\end{equation*}
and $\bmD={\mathlarger\epstensor}(\nhat)\bmE$ and $\bmE=-\nabla U$
give
\begin{equation*}
  - \frac12 \int_\Omega ( \bmD\cdot\bmE ) \, \dV =
  - \frac12 \int_\Omega \bigl[
  \epstensor(\nhat) \nabla U \cdot \nabla U \bigr] \, \dV .
\end{equation*}
Thus $\calF$ is not the appropriate free energy for modeling dynamics
that include the coupled evolution of the electric field in the cell
or the charge dynamics of the electric circuit.  The potential energy
$\calG$, on the other hand, is valid in the absence of any of these
equilibrium or instantaneous-adjustment assumptions.


The time scales for the various dynamic processes in our system vary
widely and depend on details of any specific experiment being modeled.
At one end of the spectrum is the switch-off time ($\tauK$ in our
notation, the slowest time scale, of the order of $10^{-2}$\,s).  At
the other end is the time scale for the evolution of the electric
field in the cell, $\tauE$, governed by the time-dependent Maxwell
equations, which is of the order of $10^{-13}$\,s for the kinds of
systems of interest to us.  In between these extremes are the
switch-on time, $\tauV$, and the time scale for the dynamics of the
electric circuit, $\tausig$.  The switch-on time $\tauV$ is
proportional to $1/V^2$ (where $V$ is the applied voltage) and can
vary from values comparable to $\tauK$ to values of the order of
$10^{-8}$\,s.  There is some overlap between values that $\tauV$ can
take and those that $\tausig$ can take, and our modeling has been
concerned with such situations.  The evolution of the electric field
in the cell is several orders of magnitude faster than any of these,
and its response has been treated as instantaneous, as is always
assumed in modeling such systems.

For the example problem that we have analyzed in detail (a
splay-\Freed\ cell), the final form of the dynamical system
\eqref{eqn:coupled-dynamical-system} is quite clean, with everything
given in simple explicit analytical expressions.  This is, of course,
a consequence of our modeling assumptions.  The assumption that fields
in the cell depend on only one space variable buys one a lot.  It
gives $D_z=\text{const}$ and $\sigma=\text{const}$ (in equilibrium) or
$\sigma=\sigma(t)$ (in dynamics), and it enables us to write an
explicit analytical expression for the capacitance of the cell,
$C[\nhat]$ in \eqref{eqn:Cn}, and for the electric field in the cell,
as in \eqref{eqn:EUzDzsigma} and \eqref{eqn:Uz}.  If one abandons this
assumption and allows fields in the cell to depend on more than one
space variable, then all these consequences are lost.  The assumption
of fields depending on one space dimension is very common (and
appropriate) in modeling thin-film liquid-crystal systems.

The interplay between electric fields and liquid crystals has been of
interest and importance ever since the discovery of the electro-optic
effect in the 1970s---for an early review, see \cite{goodman:75} and
references therein.  Bringing the electric circuit into the picture
(as we have done here) has illuminated the role of the battery in
producing the term
$-\frac12{\mathlarger\epstensor}(\nhat)\nabla U \cdot \nabla U$ in the
free energy \eqref{eqn:FnU}, and it has highlighted what assumptions
must be made to express the free energy in that form.  If one merely
wanted to obtain a coupled dynamical system such as
\eqref{eqn:coupled-dynamical-system}, then one could have proceeded
more directly, starting with appropriate equations for director
dynamics and circuit dynamics
\begin{equation*}
  \gamma_1 \frac{\partial\nhat}{\partial t} =
  \div \Bigl( \frac{\partial\We}{\partial\nabla\nhat} \Bigr) -
  \frac{\partial\We}{\partial\nhat} + \lambda \nhat +
  \eps_0 \epsa (\bmE\cdot\nhat) \bmE , \quad
  R \frac{\dQ}{\dt} + \frac1C Q = V
\end{equation*}
and coupling them via appropriate expressions for the electric field
(which depends on $Q=\sigma A$, as in \eqref{eqn:EUzDzsigma} and
\eqref{eqn:Uz}) and the capacitance (which depends on $\nhat$, as in
\eqref{eqn:Cn}).
Our more elaborated approach was motivated by a desire to see the
``full picture'' in terms of the potential energies, where they come
from, and the assumptions needed for various simplifications and
reductions.


\section*{Acknowledgments}

The impetus for this analysis came from exchanges with John Ball and
Nigel Mottram, and we are grateful to both of them for their thoughts,
suggestions, and feedback on versions of this note.  We are also
grateful to Lev Truskinovsky for discussions much earlier that planted
the seeds for some of these ideas, and for the reference
\cite{fosdick:truskinovsky:03}, which illuminates some related
concepts.


\bibliography{paper}

\end{document}